\documentclass[oldversion]{aa}
\usepackage{epsfig}
\usepackage{natbib}
\usepackage{lscape}
\usepackage{wasysym}
\usepackage{amssymb}
\usepackage{longtable}
\usepackage{rotating}
\usepackage{color}
\usepackage{colortbl}
\usepackage{fancyheadings}
\bibpunct{(}{)}{;}{a}{}{,}

\newcommand\T{\rule{0pt}{2.6ex}}
\newcommand\B{\rule[-1.2ex]{0pt}{0pt}}

\begin{document}

\title{Chemical abundances of 451 stars from the HARPS GTO planet search program \thanks{Based on observations collected at the La Silla Parana Observatory, ESO (Chile) with the HARPS spectrograph at the 3.6m telescope, under the GTO program 072.C-0488.}}
\subtitle{Thin disc, thick disc, and planets}

\author{ V. Neves\inst{1,2} \and N. C. Santos\inst{1,3} \and S. G. Sousa\inst{1,4} \and A. C. M. Correia\inst{2} \and G. Israelian\inst{5} } 

\institute{
    Centro de Astrof{\'\i}sica, Universidade do Porto, Rua das Estrelas,
    4150-762 Porto, Portugal
    \and
    Departamento de F{\'\i}sica, Universidade de Aveiro, Campus de Santiago
    3810-193 Aveiro, Portugal
    \and
    Observatoire de Gen\`eve, 51 ch. des Maillettes, 1290 Sauverny, Switzerland
    \and
    Departamento de Matem\'{a}tica Aplicada, Faculdade de Ci\^{e}ncias da Universidade do Porto, Portugal
    \and
    Instituto de Astrof{\'\i}sica de Canarias, 38200 La Laguna, Tenerife, Spain
}


\date{Received XXX; accepted XXX}

\abstract{ We present a uniform study of the chemical abundances of 12 elements (Si, Ca, Sc, Ti, V, Cr, Mn, Co, Ni, Na, Mg, and Al) derived from the spectra of 451 stars observed as part of one of the HARPS GTO planet search programs. {Sixty eight} of these are planet-bearing stars. The main goals of our work are: i) the investigation of possible differences between the abundances of stars with and without planets; ii) the study of the possible differences in the abundances of stars in the thin and the thick disc. We confirm that there is a {systematically higher metallicity in planet host stars, when compared to non planet-hosts,} common to all studied species. We also found that there is no difference in the galactic chemical evolution trends of the stars with and without planets. 
Stars that harbour planetary companions simply appear to be in the high metallicity tail of the distribution. We also confirm that Neptunian and super-Earth class planets may be easier to find at lower metallicities. A statistically significative abundance difference between stars of the thin and the thick disc was found for [Fe/H] $<$ 0. However, the populations from the thick and the thin disc cannot be clearly separated. 

	    }

\authorrunning{Neves et al.}
\titlerunning{Chemical abundances of 451 stars}
\maketitle

\section{Introduction}

The discovery that giant-planet host stars have, on average, a higher metallicity than non-planet hosts \citep{Gonzalez-1997}, inspired a series of spectroscopic studies for stars studied as part of radial-velocity planet-hunting programs 
\citep[e.g.][]{Santos-2000b,Santos-2001a,Santos-2004b,Laws-2003,Fischer-2005}. Gradually, similar abundance studies were made for elements other than iron \citep[e.g.][]{Gonzalez-2001,Sadakane-2002,Israelian-2003,Bodaghee-2003,Beirao-2005,Fischer-2005,Gilli-2006,Ecuvillon-2006b,Takeda-2007}. The same abundance difference was observed in these elements, suggesting that the probability of finding a giant planet is strongly related to the abundance content of its host star.

Interestingly, stars with Neptune and super-Earth-class planets might have different trends than the ones found for Jovians. There are hints that these planets may be easier to find at the typical metallicities of non-planet (non-Jovian?) host stars or even at lower [Fe/H] values \citep[][]{Udry-2006,Sousa-2008}. However, there are still an insufficient number of planets for compiling a robust statistical distribution.

Some authors also referred to the existence of different [X/Fe] trends in planet-host stars compared to stars without planets for the same [Fe/H]. These potential differences were detected for some elements \citep[e.g. Si, Ti, Mn, V, Co, Ni, Na, and Mg --][]{Bodaghee-2003,Gilli-2006,Robinson-2006} and are difficult to explain with galactic {chemical-evolution} models. Despite that, the results obtained are still contradictory and no definitive conclusion has yet been reached. A review of these trends was recently published by \citet{Gonzalez-2007}.

The large amount of spectra used in planet-search surveys with radial velocity methods may also be employed to study the abundances and  kinematic properties of the stars in the solar neighbourhood \citep[e.g.][]{Santos-2003,Fischer-2005,Gilli-2006,Ecuvillon-2007}. This data can be used to study the properties of the different populations of the Galaxy and, in particular, the differences between the thin and thick disc. \citet{Fuhrmann-1998} showed that the thick disc stars have a distinctive [Mg/Fe] overabundance compared to the stars in the thin disc, for the same [Fe/H], at least for metallicities below solar. This fact was confirmed by \citet{Bensby-2003} and expanded for other elements. The work of \citet{Ecuvillon-2007}, however, did not find any clear difference in the element abundance for the different populations observed in previous studies.

In this paper, we derive the abundances of twelve species (silica, calcium, titanium, scandium, manganese, chromium, vanadium, cobalt, nickel, sodium, magnesium, and aluminium), based on spectra of the same HARPS guaranteed time observations (GTO) from the planet search program. This sample contains 451 stars, of which {68} are planet hosts and the remaining {383} stars are dwarfs with no known orbiting planets \citep{Sousa-2008}. We also derive the kinematical properties of the sample. The main goals of this work are twofold: i) the investigation of possible differences between the abundances of stars with and without planets; ii) the study of the possible differences between the abundances of stars from the thin and thick disc of the Galaxy. In Sect. \ref{sec:sample}, we present the sample used in this work, along with the calculation of the galactic space velocity data and the selection of different groups of stars, based on their kinematic properties. The method and the selection steps that precede the chemical abundance determination will be explained in Sect. \ref{sec:chemical}. Section \ref{sec:abundance} contains the abundance analysis of the full sample as well as the final [X/H] values. It also includes discussion of the uncertainties and errors in our methodology as well as a comparison of our results with the literature. A discussion of our results follows in Sect. \ref{sec:planet}, where we explore the possible differences between stars with and without planets in our sample.  
In Sect. \ref{sec:disc}, we discuss the differences between the abundances of the thin and thick disc stars, using the [X/Fe] distributions for [Fe/H] $<$ 0 and the [X/Fe] versus [Fe/H] plots. We also use these plots to complete a brief study of the trends in the Galactic chemical evolution. Finally, in Sect. \ref{sec:conclusion}, we draw our concluding remarks.

\begin{table*}[t!] 
  \centering
\caption[]{Sample table of the parameters used to assign the Galactic population to which each star belongs. Columns 2, 3, and 4 list the U,V and W velocities relative to the local standard of rest (LSR). Columns 5, 6, and 7 list the Gaussian distributions of each population: thin disc, thick disc, and halo, respectively. In Cols. 8 and 9, we provide the relative probabilities of the thick disc relative to the thin disc and of the thick disc relative to the halo. The last {column} depicts the probable population where each star belongs.}
  \label{table:disc_data}
  \begin{tabular}{ c r r r r r r r r c}
  \hline
  \hline

Star ID \T & $U_{LSR}$ & $V_{LSR}$ & $W_{LSR}$ & $f_{thin}$ & $f_{thick}$ & $f_{halo}$ & $\frac{P(thickdisk)}{P(thindisk)}$ & $\frac{P(thickdisk)}{P(halo)}$ & group \\
  \B&  \multicolumn {3}{c}{[km s$^{-1}$]} &	&	&	&	&	&	\\
\hline
... & ... & ... & ... & ... & ... & ... & ... & ... & ... \\
HD134060 &   26 &  -34 &   15 &   1.75E-06 &   5.75E-07 &   5.67E-09 &      0.04 & 6765.43 & thin \\
HD134606 &  -11 &  -16 &    1 &   5.36E-06 &   5.20E-07 &   3.80E-09 &      0.01 & 9134.99 & thin \\
HD134664 &   27 &    2 &  -14 &   1.92E-06 &   2.69E-07 &   2.26E-09 &      0.02 & 7933.64 & thin \\
HD134985 &  -29 &  -65 &  -41 &   7.14E-09 &   2.92E-07 &   9.84E-09 &      4.55 & 1979.36 & thick \\
HD134987 &  -11 &  -35 &   28 &   7.24E-07 &   4.91E-07 &   5.62E-09 &      0.08 & 5824.82 & thin \\
HD136352 & -109 &  -42 &   43 &   4.76E-10 &   8.89E-08 &   4.93E-09 &     20.78 & 1203.01 & thick \\
HD136713 &    9 &  -13 &   -5 &   5.15E-06 &   4.84E-07 &   3.51E-09 &      0.01 & 9182.53 & thin \\
HD136894 &    5 &    3 &    0 &   3.76E-06 &   3.11E-07 &   2.28E-09 &      0.01 & 9078.65 & thin \\
... & ... & ... & ... & ... & ... & ... & ... & ... & ... \\
\hline
\end{tabular}
\end{table*}

\section{The sample}
\label{sec:sample}

The spectra for the HARPS ``high precision'' GTO program were obtained with the HARPS spectrograph at the ESO 3.6m telescope (La Silla, Chile). This {sub-sample} consists of 451 stars, selected from the volume-limited sample of solar neighbourhood stars studied with the CORALIE spectrograph \citep{Udry-2000} as well as a group of planet hosts from the southern hemisphere. Details about this sample are given in \citet{Sousa-2008}. We note that we updated the original GTO catalogue using data from the extrasolar planets encyclopaedia\footnote{http://exoplanet.eu/}, to take account of the newly discovered planets orbiting the stars HD40307 \citep{Mayor-2008}, HD47186, and HD181433 (Bouchy et al. in prep.). The total number of planet-bearing stars in the sample is now {68}. Most stars in the catalogue are slow rotators, non-evolved, and with a low chromospheric activity level. 

The individual spectra were reduced using the HARPS pipeline and later combined with IRAF\footnote{IRAF is distributed by National Optical Astronomy Observatories, operated by the Association of Universities for Research in Astronomy, Inc., under contract with the National Science Foundation, U.S.A.} after correcting for the radial velocities of the stars. The final spectra have a resolution of {$R\sim110\,000$} and a signal-to-noise ratio ranging from $\sim70$ to $\sim2000$ depending on the amount and quality of the original spectra. Nine tenths of the spectra have S/N higher than 200 and 50\% have a S/N higher than 450.

A solar spectrum was also collected with HARPS using solar light reflected by the asteroid Ceres. This spectrum has the same spectral resolution as the remaining spectra and a S/N ratio of $\sim250$. 

The stellar parameters used in the present paper were determined by \citet{Sousa-2008}. We refer to this paper for details.

\subsection {The UVW data and thin/thick disc stars}
\label{sec:uvwdata}

The galactic space-velocity components (UVW), analysed in Sect. \ref{sec:disc}, were computed with the trigonometric parallaxes and proper motions from the Hipparcos catalogue \citep{ESA-1997}, and by adopting the mean radial velocities obtained from the HARPS spectra (courtesy of the HARPS GTO team).  The Galactic space-velocity components were calculated using the procedure from \citet{Johnson-1987} and corrected for the solar motion relative to the local standard of rest (LSR) using $(U_\odot,V_\odot,W_\odot)=(+10.00,+5.25,+7.17)$ km s$^{-1}$ from \citet{Dehnen-1998}. This LSR correction was also used by \citet{Bensby-2003,Bensby-2005}.

The selection of the thin and thick disc stars was completed using the method proposed by \citet[][]{Bensby-2003,Bensby-2005}. The characteristic velocity dispersions ($\sigma_U,\sigma_V, \sigma_W$), the asymmetric drift, $V_{asym}$, and the observed fractions of these populations in the solar neighbourhood were taken from \citet{Bensby-2005}, as well as the equations to calculate the relative probability of a star belonging to a certain population. Each probability (P) is calculated by multiplying its Gaussian probability distribution with the observed fraction of the respective population. Following the recommendations of \citet{Bensby-2005}, if $P_{thick disc}/P_{thin disc}\geq2$, the star will be considered a member of the thick disc. Otherwise, if $P_{thick disc}/P_{thin disc}\leq0.6$,  the star will belong to the thin disc. All remaining stars having a probability between these values will be included in a transition population.

The star HD104006 with a $P_{thick disc}/P_{halo}=0.286$ will be considered as belonging to the halo. 

\begin{figure}[t]
\centering
\includegraphics[width= 9 cm]{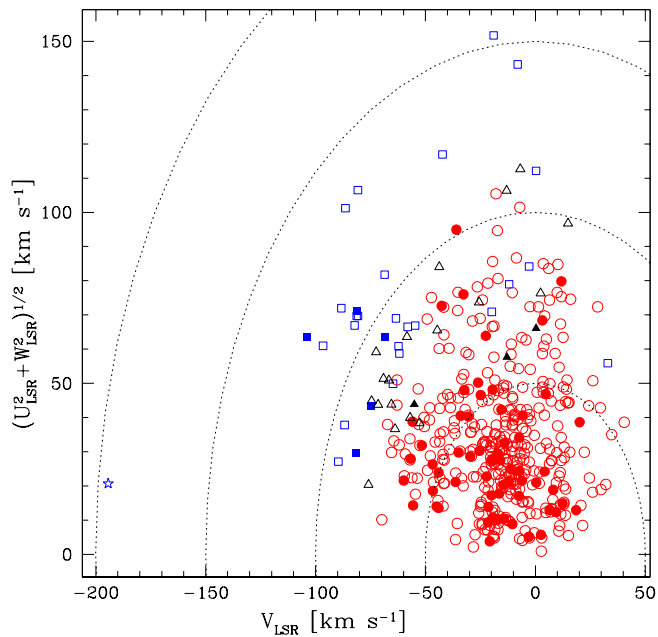}
\caption[]{Toomre diagram for the full sample (451 stars). The stars from the thin disc, transition group, and the thick disc are marked by red circles, black triangles, and blue squares, respectively. The stars with and without planets are shown with full and empty symbols, respectively.The halo star HD104006 is represented by a starred symbol.}
\label{fig:tommre}
\end{figure}

In Fig. \ref{fig:tommre}, the Toomre Diagram shown represents the entire sample. The blue squares, black triangles, and red circles correspond to the thick disc, transition and thin disc stars, respectively. The stars with and without planets are indicated by filled and empty symbols, respectively. The halo star HD104006 is represented by a starred symbol. A sample of the distributions and probabilities calculated for each star, as well as the parameters used in their calculation, are presented in Table \ref{table:disc_data}. Among the 451 stars in our sample, we have 400 stars from the thin disc, 29 from the thick disc, and 21 are considered to be transition stars that do not belong to any group or have a determined origin. According to this adopted criteria, only 1 star is from the halo.

\section{Chemical abundances}
\label{sec:chemical}

The chemical abundances of the studied elements were derived using a differential local thermodynamic equilibrium (LTE) analysis. This was completed with the 2002 version of the MOOG\footnote{The source code of MOOG2002 can be downloaded at http://verdi.as.utexas.edu/moog.html} program \citep{Sneden-1973} using the `abfind' driver. A grid of \citet{Kurucz-1993} ATLAS9 atmospheres were used as input, along with the equivalent widths and the atomic parameters, wavelength ($\lambda$), excitation energy of the lower energy level ($\chi_l$), and oscillator strength ($\log gf$) of each line. The atmospheric parameters, effective temperature ($T_{eff}$), surface gravity ($\log g$), microturbulence ($\xi_t$), and metallicity ([Fe/H])
were taken from \citet{Sousa-2008} and derived from the same spectra used in our study. This provided consistency to our work, ensuring that it was more uniform and less affected by external errors. The reference abundances used in the differential analysis are those of the Sun, and were taken from \citet{Anders-1989}.

\subsection{The line list}
\label{sec:linelist}

The initial line list, along with the atomic parameters (including an initial estimate of the oscillator strengths) were taken from the VALD\footnote{Vienna Atomic Line Database} online database \citep[][]{Piskunov-1995,Kupka-1999}. When extracting the lines, we used the solar stellar parameters as input: $T_{eff}$ = 5777 K; $\log g$ = 4.44 dex; $\xi_t$ = 1.0 km/s. 
We found 2169 lines in the spectral region of interest (4500 to 6910 \AA). After a lengthy process of selection and identification of the `good lines', involving the use of the IRAF {``splot''} tool with the Kurucz Solar Flux Atlas \citep{Kurucz-1984}, we obtained a shorter list of 364 lines. All strongly blended lines, as well as lines that were too weak (having Equivalent Widths (hereafter EW) below 5m\AA), or too strong (lines with EW $>200$ m\AA), and those inside the wings of strong lines (e.g. $H_\alpha$ ($\lambda=6562.81$ \AA), $H_\beta$ ($\lambda=4861.34$ \AA), and $\lambda5172.70,\lambda5183.62$ Mg I lines) were excluded in this process. 

\begin{table*}[!t]\scriptsize
\centering
\caption[]{Wavelength ($\lambda$), excitation potential ($\chi_l$), oscillator strength ($\log gf$) and solar equivalent widths (EW$_\odot$) for the lines selected in the present paper.}
\label {table:final_list}
  \begin{tabular}{c c r r | c c r r | c c r r}
\hline
\hline
$\lambda$ \T & $\chi_l$ & $\log gf$ & EW$_\odot$  &$\lambda $ & $\chi_l$ & $\log gf$ & EW$_\odot$ &$\lambda $ & $\chi_l$ & $\log gf$ & EW$_\odot$ \\
$[\AA]$ \B & [eV] &  & [m\AA{}] & [\AA] & [eV] &  & [m\AA{}] & [\AA] & [eV] &  & [m\AA{}] \\
\hline

\multicolumn{3}{l}{\textbf{Na I - 2 lines}, $\log\epsilon_\circ=6.33$} & & 5039.96 & 0.02 & -1.199 &  75.3 & 5737.07 & 1.06 & -0.815 &  10.4\\
5688.22 & 2.10 & -0.628 & 121.4 & 5064.06 & 2.69 & -0.471 &   5.5 & 6081.45 & 1.05 & -0.692 &  14.1\\
6154.23 & 2.10 & -1.622 &  36.6 & 5071.49 & 1.46 & -0.797 &  28.8 & 6224.51 & 0.29 & -1.935 &   5.3\\
6160.75 & 2.10 & -1.363 &  54.3 & 5113.44 & 1.44 & -0.861 &  27.0 & 6251.83 & 0.29 & -1.431 &  15.0\\
\multicolumn{3}{l}{\textbf{Mg I - 3 lines}, $\log\epsilon_\circ=$ 7.58} & & 5145.47 & 1.46 & -0.622 &  37.0 & 6274.66 & 0.27 & -1.751 &   8.3\\
4730.04 & 4.35 & -2.234 &  69.6 & 5219.70 & 0.02 & -2.254 &  28.1 & 6285.17 & 0.28 & -1.676 &   9.5\\
5711.09 & 4.35 & -1.777 & 105.6 & 5490.16 & 1.46 & -1.008 &  21.4 & \multicolumn{3}{l}{\textbf{Co I - 9 lines}, $\log\epsilon_\circ=$ 4.92} & \\
6319.24 & 5.11 & -2.300 &  25.2 & 5503.90 & 2.58 & -0.218 &  12.3 & 4594.63 & 3.63 & -0.279 &  12.5\\
\multicolumn{3}{l}{\textbf{Al I - 2 lines}, $\log\epsilon_\circ=$ 6.47} & & 5648.57 & 2.49 & -0.410 &  10.1 & 4792.86 & 3.25 & -0.080 &  32.7\\
6696.03 & 3.14 & -1.571 &  36.2 & 5662.16 & 2.32 & -0.123 &  23.5 & 4813.48 & 3.22 &  0.177 &  45.9\\
6698.67 & 3.14 & -1.886 &  21.1 & 5739.48 & 2.25 & -0.781 &   7.7 & 5301.05 & 1.71 & -1.950 &  19.5\\
\multicolumn{3}{l}{\textbf{Si I - 18 lines}, $\log\epsilon_\circ=$ 7.55} & & 5766.33 & 3.29 &  0.326 &   9.6 & 5342.71 & 4.02 &  0.606 &  32.3\\
5517.54 & 5.08 & -2.496 &  12.9 & 5965.84 & 1.88 & -0.492 &  26.7 & 5352.05 & 3.58 &  0.004 &  24.4\\
5645.61 & 4.93 & -2.068 &  35.8 & 5978.55 & 1.87 & -0.602 &  22.6 & 5359.20 & 4.15 &  0.040 &   9.6\\
5684.49 & 4.95 & -1.642 &  61.2 & 6064.63 & 1.05 & -1.941 &   8.3 & 5647.24 & 2.28 & -1.594 &  14.0\\
5701.11 & 4.93 & -2.034 &  37.7 & 6091.18 & 2.27 & -0.445 &  15.0 & 6814.95 & 1.96 & -1.822 &  18.8\\
5753.64 & 5.62 & -1.333 &  45.7 & 6126.22 & 1.07 & -1.416 &  22.1 & \multicolumn{3}{l}{\textbf{Ni I - 48 lines}, $\log\epsilon_\circ=$ 6.25} & \\
5772.15 & 5.08 & -1.669 &  52.0 & 6258.11 & 1.44 & -0.435 &  51.5 & 4512.99 & 3.71 & -1.467 &  19.4\\
5797.87 & 4.95 & -1.912 &  43.9 & 6261.10 & 1.43 & -0.491 &  49.2 & 4811.99 & 3.66 & -1.363 &  25.6\\
5948.54 & 5.08 & -1.208 &  85.4 & 6599.12 & 0.90 & -2.069 &   9.3 & 4814.60 & 3.60 & -1.670 &  16.7\\
6125.02 & 5.61 & -1.555 &  31.7 & \multicolumn{3}{l}{\textbf{Ti II - 8 lines}, $\log\epsilon_\circ=$ 4.99} & & 4913.98 & 3.74 & -0.661 &  55.7\\
6142.49 & 5.62 & -1.520 &  33.1 & 4583.41 & 1.16 & -2.840 &  31.8 & 4946.04 & 3.80 & -1.224 &  26.2\\
6145.02 & 5.62 & -1.425 &  38.8 & 4636.33 & 1.16 & -3.152 &  19.8 & 4952.29 & 3.61 & -1.261 &  32.3\\
6195.46 & 5.87 & -1.666 &  17.1 & 4657.20 & 1.24 & -2.379 &  49.3 & 4976.33 & 1.68 & -3.002 &  37.7\\
6237.33 & 5.61 & -1.116 &  61.1 & 4708.67 & 1.24 & -2.392 &  48.9 & 4995.66 & 3.63 & -1.611 &  17.9\\
6243.82 & 5.62 & -1.331 &  44.8 & 4911.20 & 3.12 & -0.537 &  52.3 & 5010.94 & 3.63 & -0.901 &  48.8\\
6244.48 & 5.62 & -1.310 &  46.2 & 5211.54 & 2.59 & -1.490 &  32.8 & 5081.11 & 3.85 &  0.064 &  93.5\\
6527.21 & 5.87 & -1.227 &  38.9 & 5381.03 & 1.57 & -1.904 &  59.1 & 5094.41 & 3.83 & -1.108 &  30.3\\
6721.85 & 5.86 & -1.156 &  44.0 & 5418.77 & 1.58 & -2.104 &  49.5 & 5392.33 & 4.15 & -1.354 &  12.0\\
6741.63 & 5.98 & -1.625 &  15.5 & \multicolumn{3}{l}{\textbf{Mn I - 6 lines}, $\log\epsilon_\circ=$ 5.39} & & 5435.86 & 1.99 & -2.432 &  51.7\\
\multicolumn{3}{l}{\textbf{Ca I - 12 lines}, $\log\epsilon_\circ=$ 6.36} & & 4502.21 & 2.92 & -0.523 &  57.0 & 5462.50 & 3.85 & -0.880 &  41.0\\
5261.71 & 2.52 & -0.677 &  97.7 & 4671.77 & 2.89 & -1.567 &  14.8 & 5587.87 & 1.93 & -2.479 &  52.9\\
5349.47 & 2.71 & -0.581 &  94.6 & 4739.11 & 2.94 & -0.462 &  60.1 & 5589.36 & 3.90 & -1.148 &  26.7\\
5512.98 & 2.93 & -0.559 &  84.8 & 5377.62 & 3.84 & -0.068 &  40.8 & 5625.32 & 4.09 & -0.731 &  37.8\\
5867.56 & 2.93 & -1.592 &  25.1 & 5399.47 & 3.85 & -0.104 &  38.5 & 5628.35 & 4.09 & -1.316 &  14.7\\
6156.02 & 2.52 & -2.497 &   9.6 & 5413.67 & 3.86 & -0.476 &  21.6 & 5638.75 & 3.90 & -1.699 &   9.8\\
6161.29 & 2.52 & -1.313 &  60.6 & \multicolumn{3}{l}{\textbf{Cr I - 22 lines}, $\log\epsilon_\circ=$ 5.67} & & 5641.88 & 4.11 & -1.017 &  24.1\\
6166.44 & 2.52 & -1.155 &  69.9 & 4575.11 & 3.37 & -1.004 &  10.0 & 5643.08 & 4.16 & -1.234 &  15.1\\
6169.04 & 2.52 & -0.800 &  92.2 & 4600.75 & 1.00 & -1.457 &  82.3 & 5694.99 & 4.09 & -0.629 &  43.1\\
6449.82 & 2.52 & -0.733 &  98.1 & 4626.18 & 0.97 & -1.467 &  83.3 & 5748.36 & 1.68 & -3.279 &  28.0\\
6455.60 & 2.52 & -1.404 &  56.3 & 4633.25 & 3.13 & -1.215 &  10.4 & 5805.22 & 4.17 & -0.604 &  40.8\\
6471.67 & 2.53 & -0.825 &  91.2 & 4700.61 & 2.71 & -1.464 &  14.0 & 5847.00 & 1.68 & -3.410 &  23.0\\
6499.65 & 2.52 & -0.917 &  85.8 & 4708.02 & 3.17 & -0.104 &  55.0 & 5996.73 & 4.24 & -1.010 &  20.3\\
\multicolumn{3}{l}{\textbf{Sc I - 3 lines}, $\log\epsilon_\circ=$ 3.10} & & 4730.72 & 3.08 & -0.345 &  46.3 & 6086.29 & 4.27 & -0.471 &  43.5\\
4743.82 & 1.45 &  0.297 &   8.0 & 4767.86 & 3.56 & -0.599 &  16.1 & 6108.12 & 1.68 & -2.512 &  65.0\\
5520.50 & 1.87 &  0.562 &   6.6 & 4775.14 & 3.55 & -1.025 &   6.9 & 6111.08 & 4.09 & -0.823 &  34.2\\
5671.82 & 1.45 &  0.533 &  14.6 & 4936.34 & 3.11 & -0.343 &  45.6 & 6119.76 & 4.27 & -1.316 &  10.9\\
\multicolumn{3}{l}{\textbf{Sc II - 6 lines}, $\log\epsilon_\circ=$ 3.10} & & 4964.93 & 0.94 & -2.577 &  38.1 & 6128.98 & 1.68 & -3.368 &  25.3\\
5526.82 & 1.77 &  0.140 &  76.3 & 5214.14 & 3.37 & -0.784 &  16.4 & 6130.14 & 4.27 & -0.938 &  22.1\\
5657.88 & 1.51 & -0.326 &  66.9 & 5238.97 & 2.71 & -1.427 &  15.9 & 6175.37 & 4.09 & -0.534 &  49.0\\
5667.14 & 1.50 & -1.025 &  33.9 & 5247.57 & 0.96 & -1.618 &  82.0 & 6176.82 & 4.09 & -0.266 &  63.7\\
5684.19 & 1.51 & -0.946 &  37.2 & 5287.18 & 3.44 & -0.954 &  10.4 & 6177.25 & 1.83 & -3.538 &  14.6\\
6245.62 & 1.51 & -1.022 &  34.9 & 5296.70 & 0.98 & -1.373 &  93.1 & 6186.72 & 4.11 & -0.888 &  30.5\\
6320.84 & 1.50 & -1.863 &   8.1 & 5300.75 & 0.98 & -2.125 &  58.9 & 6204.61 & 4.09 & -1.112 &  22.0\\
\multicolumn{3}{l}{\textbf{Ti I - 30 lines}, $\log\epsilon_\circ=$ 4.99} & & 5781.18 & 3.32 & -0.886 &  15.5 & 6223.99 & 4.11 & -0.954 &  27.7\\
4555.49 & 0.85 & -0.575 &  63.7 & 5783.07 & 3.32 & -0.472 &  31.4 & 6230.10 & 4.11 & -1.132 &  20.6\\
4562.63 & 0.02 & -2.718 &  10.9 & 5787.92 & 3.32 & -0.183 &  46.0 & 6322.17 & 4.15 & -1.164 &  18.4\\
4645.19 & 1.73 & -0.666 &  22.0 & 6661.08 & 4.19 & -0.234 &  12.0 & 6327.60 & 1.68 & -3.086 &  38.6\\
4656.47 & 0.00 & -1.308 &  68.6 & 6882.52 & 3.44 & -0.392 &  32.2 & 6360.81 & 4.17 & -1.145 &  18.5\\
4675.11 & 1.07 & -0.939 &  38.1 & \multicolumn{3}{l}{\textbf{Cr II - 3 lines}, $\log\epsilon_\circ=$ 5.67} & & 6378.26 & 4.15 & -0.830 &  31.8\\
4722.61 & 1.05 & -1.433 &  18.7 & 4588.20 & 4.07 & -0.752 &  68.5 & 6598.60 & 4.24 & -0.914 &  24.9\\
4820.41 & 1.50 & -0.429 &  43.1 & 4592.05 & 4.07 & -1.252 &  47.6 & 6635.13 & 4.42 & -0.779 &  23.6\\
4913.62 & 1.87 &  0.068 &  50.2 & 4884.61 & 3.86 & -2.069 &  23.4 & 6767.78 & 1.83 & -2.136 &  79.2\\
4997.10 & 0.00 & -2.174 &  31.6 & \multicolumn{3}{l}{\textbf{V I - 7 lines}, $\log\epsilon_\circ=$ 4.00} & & 6772.32 & 3.66 & -0.963 &  49.2\\
5016.17 & 0.85 & -0.657 &  63.2 & 5670.85 & 1.08 & -0.482 &  18.8 & 6842.04 & 3.66 & -1.496 &  24.2\\
\hline
\end{tabular}
\end {table*}

The next step involved the automatic measurement of the equivalent width of the lines, in a resolution-degraded spectrum {($R\sim$110\,000)} of the Kurucz Solar Flux Atlas. This was completed with the ARES\footnote{The Automatic Routine for Line Equivalent Widths is available at: http://www.astro.up.pt/$\sim$sousasag/ares/} program \citep{Sousa-2007}. Here, we also devised a selection procedure that involved the comparison of the EWs measured in the aforementioned spectrum with those measured in the solar reflected light spectrum of the Ceres asteroid, obtained with HARPS. Both spectra have a resolution of the order of {$\sim$110\,000}. If the relative difference between the EWs, for a certain line, measured in the two spectra, was higher than 10\%, the line would be discarded. Additionally, we also excluded all lines that ARES could not reproduce well due to a poor evaluation of the lines in the spectral area of interest, or due to a poor determination of the continuum placement. These criteria are important because the abundance and especially the oscillator strength determination are very sensitive to variations in the EW. The parameters adopted in ARES were: smoother = 4, space = 2, lineresol = 0.07, miniline = 5, and rejt = 0.996 \citep[from][]{Sousa-2008}. At the end of this step, we finished with 284 lines.

\begin{figure}[t]
\includegraphics[width=9 cm]{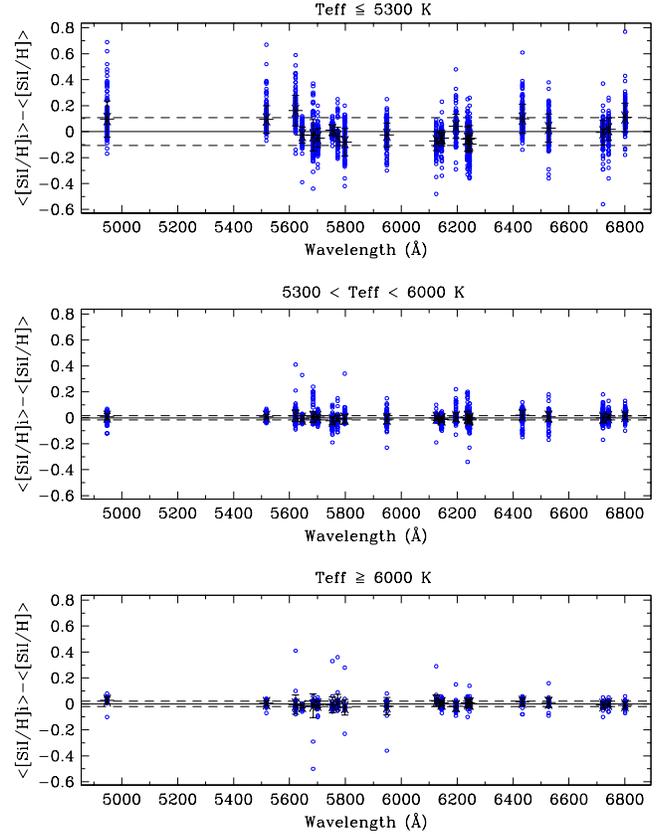}
\caption{Process of selection of the stable lines for the neutral atom of Si, shown as an example. The black crosses on the three plots represent the difference between the mean abundance of each line (from all the stars) and the mean abundance of the element. The error bars represent the rms of the mean values. The dashed lines mark the exclusion value. The open circles correspond to the difference values of the cool group ($T_{eff} \leq$ 5300 - top plot), the solar group (5300 $<T_{eff}<$ 6000 - middle plot), and the hot group ($T_{eff} \geq$ 6000 - bottom plot).}
\label{fig:line_select}
\end{figure}

The EW of these lines, along with a model atmosphere based on the solar parameters, $T_{eff}=5777$, $\log g=4.44$, $\log\epsilon_{Fe}=7.47$, and $\xi_t=1.0$, were used in the calculation of the semi-empirical oscillator strengths by using an inverse analysis with the MOOG driver {``ewfind''}. All input parameters for this driver were taken following \citet{Santos-2004b}. The new oscillator values are important because the $\log gf$ data from VALD might not be sufficiently accurate.

To test the calculated $\log gf$ values, we inverted the previous analysis using a solar atmosphere model and the measured EWs of the solar reflected light spectrum of Ceres using the {``abfind''} driver within MOOG. In general, the average abundance values found were within 0.01 dex of the expected value, except for Al ($\Delta A=\pm0.04$ with 2 lines), ScI ($\Delta A=\pm0.03$ with 3 lines) and V ($\Delta A=\pm0.02$ with 14 lines), where $A$ is the abundance of a species.

The final selection was the most important one. After determining the abundances for all stars, using the \citet{Kurucz-1993} model atmospheres calculated with the stellar parameters and EWs measured with ARES for each star and element, we divided the stars into three distinct groups: a) the cool group, containing 152 stars with $T_{eff} \leq$ 5300; b) the solar group, having 257 stars with 5300 $<T_{eff}<$ 6000; and c) the hot group, comprising 42 stars with $T_{eff} \geq$ 6000. This was completed to account for the differences in the local continuum about each line that are temperature dependent \citep{Sousa-2008}. The ARES parameters were identical to those of the solar case apart from `rejt' which depended on the S/N ratio of each spectra according to \citet{Sousa-2008}, and `miniline=2' (ensuring that all lines measured in the solar spectrum were accounted for in the other stars). Then, we calculated the difference between the mean abundance of each line (from all the stars) and the mean abundance calculated using all the lines. This was completed for each element, as we can see for Si in Fig. \ref{fig:line_select}. The open circles in the top, middle and bottom panels, represent data for hot, solar, and cool group stars, respectively, corresponding to this difference. The mean value of each line is depicted by a black cross and the error bars represent its rms.

We considered all lines with an overall line dispersion in any plot of more than a factor of 1.5 higher than the rms, to be non-stable and these lines were excluded. The dashed lines on the plots of Fig.\,\ref{fig:line_select} indicate this exclusion value. We also excluded any line with a dispersion in any plot clearly beyond the average abundance value. This selection was done with a careful manual inspection line by line and element by element. In a few cases, we chose to relax these rules slightly, especially when there was a line close to the exclusion values occurring in only one group. {We also verified each individual line for stray points clearly outside of the $2\sigma$ distribution (due to bad pixels, cosmic rays or other unknown effects). These could modify the true value of the mean or the dispersion of a certain line and lead to a wrong exclusion}. After this last selection, the final list contained 180 lines, as shown in Table \ref{table:final_list}.

\section{Abundance analysis of the full sample}
\label{sec:abundance}

As stated before, the chemical abundances of the elements were derived by completing a differential LTE analysis relative to the Sun. The analysis of the full 451 star sample was described in Sect. \ref{sec:linelist}. The final abundance for each star and element was calculated to be the average value of the abundances for all lines detected in a given star and element. A set of FORTRAN programs were developed to coordinate these calculations and compile an organised and easy-to-use database for the output data. A {sample} of our results are presented in Table \ref{table:abundances}. The complete results are available online.

\begin{table*}[t] 
\centering
\caption[Sample table of the abundance derived for the studied elements]{Sample table of the derived abundances of the elements, rms and number of measured lines (n) for each star.}
  \label{table:abundances}

 \begin{tabular}{ c | r c c | r c c | r c c | r c c c }

  \hline
  \hline
Star ID \T & [Si/H] & rms & n & [Ca/H] & rms & n & [ScI/H] & rms & n & [ScII/H] & rms & n & ... \B \\
\hline
... & ... & ... & ... & ... & ... & ... & ... & ... & ... & ... & ... & ... & ...\\
HD12617 &  0.15 & 0.10 & 17 &  0.05 & 0.14 & 10 &  0.48 & 0.31 &  3 &  0.14 & 0.11 &  6 & ...\\
HD13060 &  0.06 & 0.07 & 16 &  0.03 & 0.06 & 12 &  0.12 & 0.11 &  3 & -0.01 & 0.05 &  6 & ...\\
HD13724 &  0.24 & 0.02 & 17 &  0.22 & 0.02 & 11 &  0.30 & 0.02 &  3 &  0.31 & 0.07 &  6 & ...\\
HD13789 & -0.04 & 0.08 & 16 &  0.01 & 0.12 & 11 &  0.31 & 0.25 &  3 & -0.13 & 0.13 &  6 & ...\\
HD13808 & -0.14 & 0.07 & 16 & -0.20 & 0.11 & 11 & -0.05 & 0.11 &  3 & -0.22 & 0.08 &  6 & ...\\
HD14374 & -0.02 & 0.02 & 16 & -0.03 & 0.04 & 11 &  0.00 & 0.05 &  3 & -0.07 & 0.07 &  6 & ...\\
HD14635 &  0.02 & 0.10 & 18 &  0.05 & 0.14 & 11 &  0.49 & 0.30 &  3 & -0.04 & 0.11 &  6 & ...\\
HD14680 & -0.16 & 0.05 & 17 & -0.12 & 0.11 & 12 & -0.01 & 0.12 &  3 & -0.21 & 0.07 &  6 & ...\\
HD14744 & -0.11 & 0.04 & 18 & -0.13 & 0.06 & 11 &  0.01 & 0.17 &  3 & -0.15 & 0.08 &  6 & ...\\
HD14747 & -0.24 & 0.01 & 17 & -0.23 & 0.01 & 11 & -0.22 & 0.03 &  3 & -0.22 & 0.04 &  6 & ...\\
... & ... & ... & ... & ... & ... & ... & ... & ... & ... & ... & ... & ... & ...\\
\hline

\end{tabular}

\end{table*}

\subsection{Uncertainties and errors}
\label{sec:errors}

Uncertainties can introduce different errors in the calculated abundances. Random errors can affect the measurement of EWs for individual lines (e.g. due to noise and cosmic-ray hits) and systematic errors can occur, for example, due to blending, NLTE effects, or to a poor choice of continuum location. To minimise both types of errors, we should use high quality data and as many lines as possible for each element. Unfortunately, we were only able to select 2 or 3 lines for Na, Mg, Al, ScI and CrII. The presence of insufficient data can produce higher internal random errors, and might also introduce systematic bias that can be extremely difficult to identify in our results. Therefore, all conclusions regarding these elements should be considered with caution. We further note that we performed a uniform analysis of the sample, which minimised possible errors due to, for example, differences in line lists, atmospheric parameters, oscillator strength biases, and applied methodologies.

\subsubsection{Random errors}
\label{sec:random}

The typical rms uncertainties for the atmospheric parameters are 24 K for $T_{eff}$, 0.04 dex for $\log g$, 0.02 dex for [Fe/H], and 0.03 km s$^{-1}$ for $\xi_t$. {These uncertainties are taken from \citet{Sousa-2008} and characterize the relative precision of the derived parameters.}

We tested the sensitivity of the derived abundances to the uncertainties in the atmospheric parameters. First, for each parameter, we chose three different stars with similar atmospheric parameters except for the one being tested. Afterwards, we generated new atmospheric models, changing the tested parameter by an amount required to assure that the sensitivity of the abundances could be clearly observed ($\Delta T_{eff}=\pm$100 K, $\Delta\log g=\pm0.3$ dex, $\Delta \xi_t=\pm0.5$ km s$^{-1}$, and $\Delta$[Fe/H] $=\pm$0.3 dex). The obtained results are displayed in Table \ref{table:errors}.   

In general, we observe that the neutral elements are sensitive to temperature changes, whereas the ions are most sensitive to changes in surface gravity. The ions are also more sensitive to metallicity changes than the neutral elements, although the sensitivity is not as significant as the one for either $T_{eff}$ and $\log g$. 

\begin{table*}[t!] 
\caption[]{Abundance sensitivities of the studied elements to changes in the stellar parameters (temperature with $\Delta T_{eff}=\pm100$ K, surface gravity with $\Delta\log g=\pm0.3$ dex, metallicity with $\Delta$[Fe/H] $=\pm0.3$ dex, and microturbulence with $\Delta \xi_t=\pm0.5$ km s$^{-1}$).}
  \label{table:errors}
  \begin{tabular}{ l c c r c c c c c c c c}

  \hline
  \hline
Star ID \T & $T_{eff}$ [K] & $\log g$ & [Fe/H] & $\xi_t$ [km s$^{-1}$] & Na & Mg & Al & Si & Ca & ScI & ScII\\
\hline
\multicolumn {12}{c}{Temperature with $\Delta T_{eff}=\pm100$ K} \\
HD10567 & 4748 & 4.42 & -0.02 & 0.90 & $\pm$0.10 & $\pm$0.02 & $\pm$0.07 & $\mp$0.06 & $\pm$0.12 & $\pm$0.14 & $\mp$0.02 \\
HD20616 & 5519 & 4.43 &  0.01 & 0.94 & $\pm$0.06 & $\pm$0.06 & $\pm$0.05 & $\mp$0.01 & $\pm$0.08 & $\pm$0.10 & $\mp$0.01 \\
HD20945 & 6118 & 4.50 &  0.03 & 1.21 & $\pm$0.06 & $\pm$0.06 & $\pm$0.05 & $\pm$0.03 & $\pm$0.07 & $\pm$0.09 & $\pm$0.01 \\
\multicolumn {12}{c}{Surface gravity with $\Delta\log g=\pm0.3$ dex} \\
HD19556 & 5676 & 4.03 &  0.06 & 1.11 & $\mp$0.05 & $\mp$0.05 & $\mp$0.02 & $\pm$0.01 & $\mp$0.05 & $\mp$0.01 & $\pm$0.13 \\
HD11720 & 5667 & 4.32 &  0.22 & 1.01 & $\mp$0.07 & $\mp$0.06 & $\mp$0.02 & $\pm$0.01 & $\mp$0.07 & $\mp$0.01 & $\pm$0.12 \\
HD20260 & 5658 & 4.49 &  0.18 & 1.02 & $\mp$0.07 & $\mp$0.06 & $\mp$0.02 & $\pm$0.01 & $\mp$0.08 & $\mp$0.01 & $\pm$0.12 \\
\multicolumn {12}{c}{Microturbulence with $\Delta \xi_t=\pm0.5$ km s$^{-1}$} \\
HD9828 & 5381 & 4.42 & -0.26 & 0.64 & $\mp$0.02 & $\mp$0.03 & $\mp$0.02 & $\mp$0.01 & $\mp$0.07 & $\mp$0.02 & $\mp$0.06 \\
HD18956 & 5726 & 4.41 & -0.24 & 0.95 & $\mp$0.02 & $\mp$0.04 & $\mp$0.01 & $\mp$0.01 & $\mp$0.08 & $\mp$0.01 & $\mp$0.07 \\
HD4444 & 5999 & 4.37 & -0.22 & 1.26 & $\mp$0.02 & $\mp$0.03 & $\mp$0.01 & $\mp$0.02 & $\mp$0.08 & $\mp$0.01 & $\mp$0.09 \\
\multicolumn {12}{c}{Metallicity with $\Delta$[Fe/H] $=\pm$0.3 dex} \\
HD7874 & 5778 & 4.46 & -0.67 & 1.03 & $\mp$0.03 & $\pm$0.00 & $\mp$0.01 & $\pm$0.01 & $\pm$0.00 & $\mp$0.01 & $\pm$0.06 \\
HD9670 & 5845 & 4.39 & -0.18 & 1.04 & $\pm$0.00 & $\pm$0.01 & $\mp$0.01 & $\pm$0.02 & $\pm$0.00 & $\mp$0.01 & $\pm$0.08 \\
HD20220 & 5757 & 4.47 &  0.29 & 1.01 & $\pm$0.04 & $\pm$0.03 & $\pm$0.01 & $\pm$0.05 & $\pm$0.03 & $\pm$0.01 & $\pm$0.11 \\



\end{tabular}

  \begin{tabular}{l c c r c c c c c c c c c}

  \hline
Star ID \T & $T_{eff}$ [K] & $\log g$ & [Fe/H] & $\xi_t$ [km s$^{-1}$]	& TiI & TiII & V & CrI & CrII & Mn & Co & Ni \B \\
\hline
\multicolumn {13}{c}{Temperature with $\Delta T_{eff}=\pm100$ K} \\
HD10567 & 4748 & 4.42 & -0.02 & 0.90 & $\pm$0.14 & $\mp$0.02 & $\pm$0.15 & $\pm$0.09 & $\mp$0.08 & $\pm$0.06 & $\pm$0.02 & $\mp$0.02 \\
HD20616 & 5519 & 4.43 &  0.01 & 0.94 & $\pm$0.12 & $\mp$0.01 & $\pm$0.12 & $\pm$0.08 & $\mp$0.04 & $\pm$0.07 & $\pm$0.06 & $\pm$0.05 \\
HD20945 & 6118 & 4.50 &  0.03 & 1.21 & $\pm$0.09 & $\pm$0.01 & $\pm$0.10 & $\pm$0.08 & $\mp$0.02 & $\pm$0.07 & $\pm$0.07 & $\pm$0.06 \\
\multicolumn {13}{c}{Surface Gravity with $\Delta\log g=\pm0.3$ dex} \\
HD19556 & 5676 & 4.03 &  0.06 & 1.11 & $\mp$0.01 & $\pm$0.12 & $\mp$0.01 & $\mp$0.02 & $\pm$0.11 & $\mp$0.02 & $\pm$0.01 & $\pm$0.01 \\
HD11720 & 5667 & 4.32 &  0.22 & 1.01 & $\mp$0.01 & $\pm$0.12 & $\mp$0.01 & $\mp$0.03 & $\pm$0.11 & $\mp$0.05 & $\pm$0.03 & $\pm$0.01 \\
HD20260 & 5658 & 4.49 &  0.18 & 1.02 & $\mp$0.02 & $\pm$0.12 & $\mp$0.01 & $\mp$0.04 & $\pm$0.11 & $\mp$0.04 & $\pm$0.02 & $\pm$0.02 \\
\multicolumn {13}{c}{Microturbulence with $\Delta \xi_t=\pm0.5$ km s$^{-1}$} \\
HD9828 & 5381 & 4.42 & -0.26 & 0.64 & $\mp$0.08 & $\mp$0.06 & $\mp$0.03 & $\mp$0.06 & $\mp$0.07 & $\mp$0.07 & $\mp$0.05 & $\mp$0.05 \\
HD18956 & 5726 & 4.41 & -0.24 & 0.95 & $\mp$0.03 & $\mp$0.08 & $\mp$0.01 & $\mp$0.06 & $\mp$0.09 & $\mp$0.04 & $\mp$0.03 & $\mp$0.03 \\
HD4444 & 5999 & 4.37 & -0.22 & 1.26 & $\mp$0.03 & $\mp$0.08 & $\mp$0.01 & $\mp$0.04 & $\mp$0.12 & $\mp$0.04 & $\mp$0.03 & $\mp$0.03 \\
\multicolumn {13}{c}{Metallicity with $\Delta$[Fe/H] $=\pm$0.3 dex} \\
HD7874 & 5778 & 4.46 & -0.67 & 1.03 & $\mp$0.01 & $\pm$0.06 & $\mp$0.01 & $\mp$0.01 & $\pm$0.04 & $\mp$0.01 & $\pm$0.00 & $\pm$0.00 \\
HD9670 & 5845 & 4.39 & -0.18 & 1.04 & $\mp$0.01 & $\pm$0.07 & $\mp$0.01 & $\mp$0.01 & $\pm$0.06 & $\pm$0.00 & $\pm$0.01 & $\pm$0.02 \\
HD20220 & 5757 & 4.47 &  0.29 & 1.01 & $\pm$0.00 & $\pm$0.11 & $\pm$0.01 & $\pm$0.02 & $\pm$0.08 & $\pm$0.03 & $\pm$0.03 & $\pm$0.05 \\

\hline

\end{tabular}

\end{table*}

\begin{figure}[t]
\centering
\includegraphics[trim=0mm 0mm 0mm 10mm, clip,width= 9 cm]{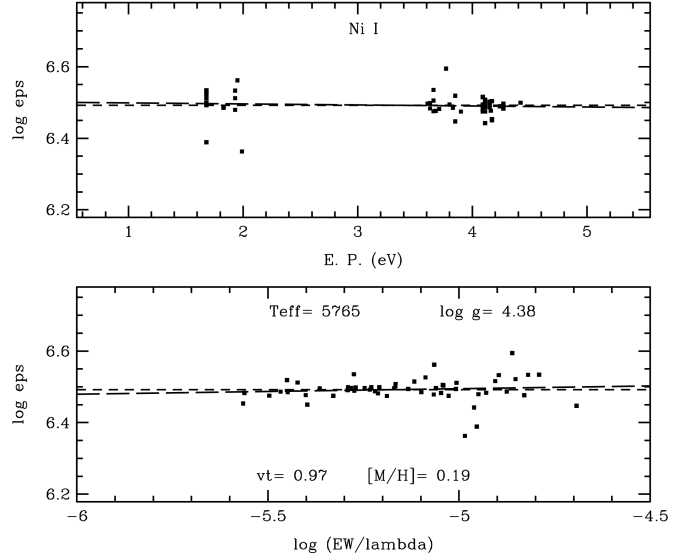}
\caption[Example of the calculation for Ni of the EP and RW slopes]{Example of the calculation of the EP and RW slopes for NiI lines for the star HD1461. The squares represent the different spectral lines, the long dashed line the calculated slope and the short dashed line the average value of the abundance ($\log$ eps). The values of the slopes are -0.010 and 0.020, respectively for the upper and lower panel.}
\label{fig:exslope}
\end{figure}

\subsubsection{Testing the stellar parameters}
\label{sec:testing}

To verify the validity of the stellar parameters, we tested our results in a variety of ways. First, we calculated the slopes of the derived abundances of the considered lines as a function of both the excitation potential (EP) and the logarithm of the reduced equivalent width (RW) of the NiI lines. In this way, we verified whether the excitation equilibrium enforced on the FeI lines of every star and used to obtain the stellar parameters \citep{Sousa-2008} was applicable to other species. In other words, we expect the slopes to be as small as possible.

We can see an example of this (star HD1461) in Fig. \ref{fig:exslope}. We have chosen nickel because its lines cover a wide range of both EP and RW, thus allowing a reliable determination of the slopes.

\begin{figure}[t]
\centering
\includegraphics[angle=-90, trim=8mm 0mm 5mm 10mm, clip,width=9 cm]{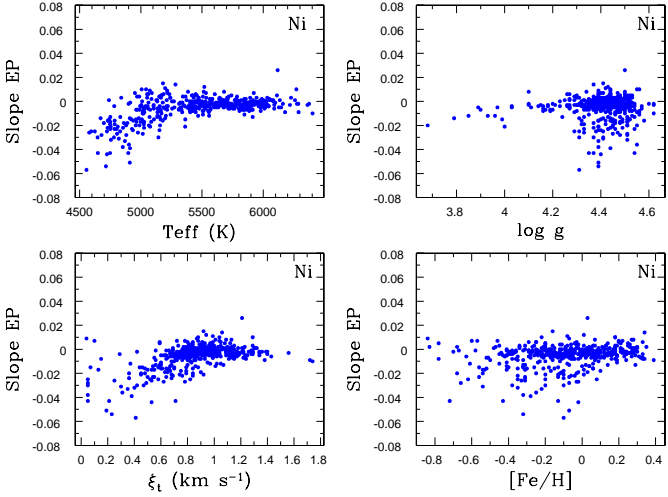}
\caption[]{EP slopes as a function of $T_{eff}$, $\log g$, $\xi_t$ and [Fe/H] for Ni.}
\label{fig:slopeEP}
\end{figure}

In Fig. \ref{fig:slopeEP}, we plot the slopes of EP, obtained for every star, as a function of the stellar parameters ($T_{eff}$, $\log g$, $\xi_t$, and [Fe/H]). From the analysis of the plots, we observe no discernible trends except in the case of the $T_{eff}$ plot, where we observe that cooler stars ($T_{eff}\lesssim 5000$ K) have a systematic bias away from the expected values. A higher dispersion is also seen, on average, for cooler stars. This might be due to the stronger line blending in cooler stars as well as to the fact that the computed $\log gf$ values (derived from the solar spectrum) may be inadequate for these stars. The same effect was noticed in the RW plots.

\begin{figure}[t]
\centering
\includegraphics[angle=-90, trim=8mm 0mm 5mm 10mm, clip, width=9cm]{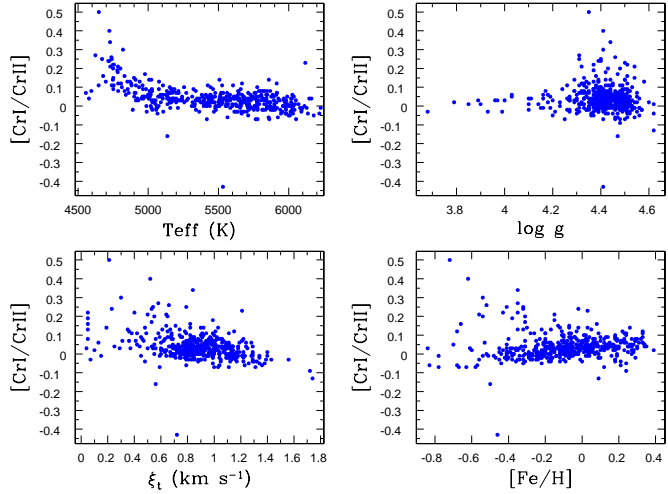}
\caption[Plots of CrI/CrII]{Plots of [CrI/CrII] as a function of $T_{eff}$, $\log g$, $\xi_t$ and [Fe/H].}
\label{fig:cr2cr1}
\end{figure}

We also plotted the [CrI/CrII] value as a function of the stellar parameters to ensure that the ionisation equilibrium enforced on the FeII lines \citep{Sousa-2008} was acceptable to other elements. 
This is depicted in Fig. \ref{fig:cr2cr1}. We expect that [CrI/CrII] should be independent of these parameters.

As seen in Fig. \ref{fig:cr2cr1}, there is a divergence in the expected ratio for stars with $T_{eff}<4900$ K. We are unable to discern any significant trend in the plots of the other parameters. This result (and that found for Ni) raises some doubt about the abundance values derived for stars with lower temperatures ($T_{eff}\lesssim4900$ K).

\begin{figure}[t]
\centering
\includegraphics[trim=0cm 1cm 0cm 1cm,clip,width=9 cm]{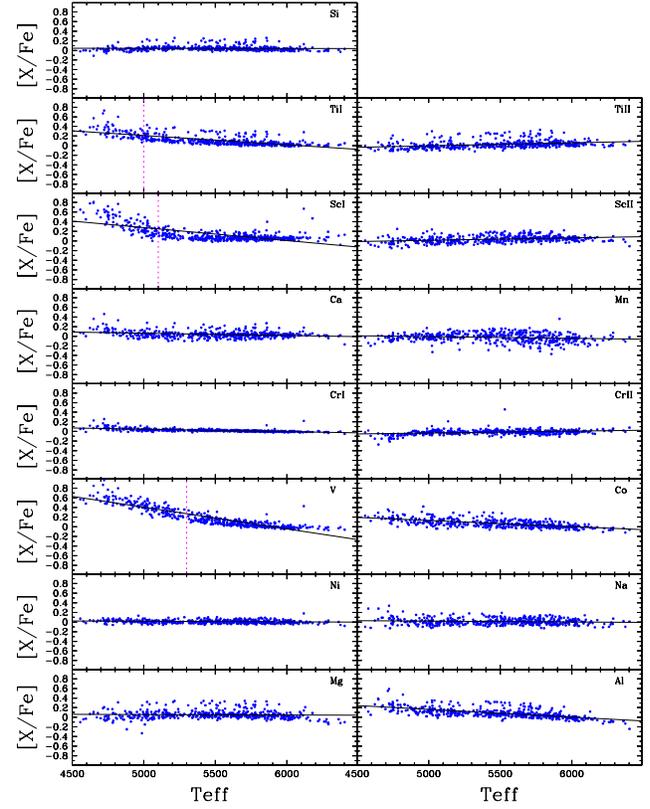}
\caption[depots]{[X/Fe] vs. $T_{eff}$ plots. The dots represent the stars of the sample. The solid lines depict the linear fits of the data. The vertical purple dotted lines indicate the cutoff temperature when appropriate.}
\label{fig:xfeteff}
\end{figure}

Finally, we plotted the abundance ratios [X/Fe] as a function of the stellar parameters. We did not detect any significant trends except for the $T_{eff}$ plot, which is presented in Fig. \ref{fig:xfeteff}. 
The slopes of [X/Fe] with $T_{eff}$ per 1000 K are listed in Table \ref{table:slopes}.

\begin{table}[b]
\centering
\caption[Slopes  per 1000 K]{Slopes of [X/Fe] ratio as a function of $T_{eff}$ per 1000K}
\begin{tabular}{ c r@{$\pm$}l | c r@{$\pm$}l }
\hline
\hline
Species \T & \multicolumn {2}{c}{Slope(T)$\pm$rms} & Species & \multicolumn {2}{c}{Slope(T)$\pm$ rms} \B \\
\hline
         Si &   -0.004  &   0.051 & CrI &   -0.053   &  0.026 \\
         TiI &    -0.192  &  0.081 & CrII &    0.039   &  0.048 \\
        TiII &    0.062    & 0.076 & Co &    -0.128   &  0.067 \\
         ScI &    -0.269  &    0.128 & Ni &   -0.012   &  0.034 \\
        ScII &    0.050  &   0.065 & Na &   -0.017   &  0.066 \\
          Ca &   -0.045   &   0.069 & Mg &  -0.009   &  0.089 \\
          Mn &   -0.034   &  0.093 & Al &    -0.161   &  0.080 \\
           V &    -0.452   &   0.086 \\
\hline

\end{tabular}
\label {table:slopes}
\end{table}

We observe significant trends with $T_{eff}$ for TiI, ScI, V, Co, and Al. It is clearly evident that the trends drift away from the expected values, except in the cases of Co and Al, are heavily influenced by the cooler stars for which the abundance might have been overestimated due to blending effects; it may also be sensitive to either deviations from excitation or ionisation equilibrium, or to problems associated with the differential analysis. We must indeed emphasise that the oscillator strengths were calculated for the Sun. To account for these effects, we decided to establish a cutoff temperature: $T_{cutoff}=5000$ K for TiI, $T_{cutoff}=5100$ K for ScI, and $T_{cutoff}=5300$ K for V. Only stars of temperature above these values are considered in the remainder of this paper, for the aforementioned elements. We note this when we analyse our results. 

\subsection{Comparison of the abundances with the literature}
\label{sec:comparison}

\begin{figure}[h]
\centering
\includegraphics[trim=0cm 1.5cm 0cm 1cm,clip,width=9 cm]{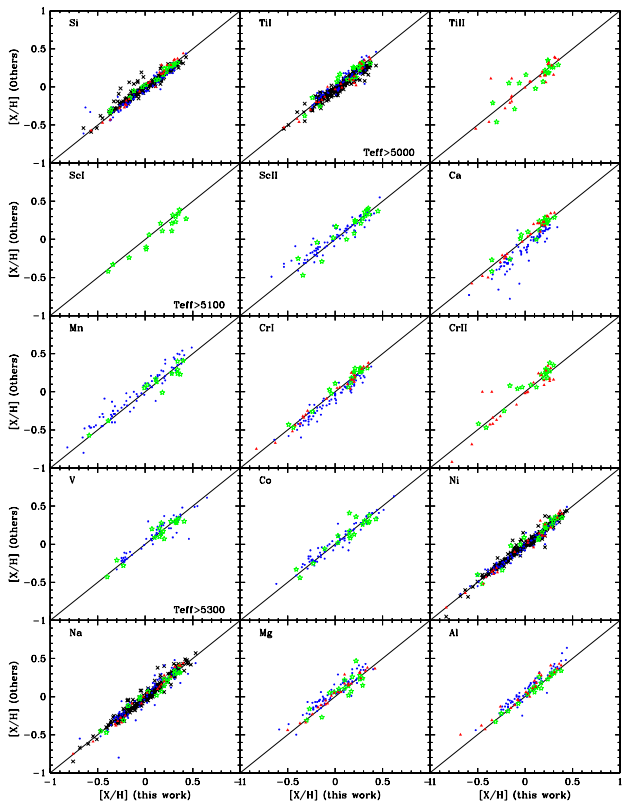}
\caption[Comparison of abundances with other authors]{Comparison of the derived abundances with those listed by other authors: \citet{Bensby-2005} (red triangles), \citet{Valenti-2005} (black crosses), \citet{Gilli-2006} (blue dots) and \citet{Takeda-2007} (green stars). The elements are identified in the top left corner of each plot.}

\label{fig:comparison}
\end{figure}

To test the reliability of our results, we compare the derived abundances with those obtained by \citet{Bensby-2005}, \citet{Valenti-2005}, \citet{Gilli-2006}, and \citet{Takeda-2007}, for stars in common, as seen in Fig.\,\ref{fig:comparison}. In this way, it was possible to verify qualitatively any systematic errors in our abundance determinations.

It is evident that, in general, our results agree with the [X/H] obtained by other authors. However, we observe a systematic underabundance trend in Ca, CrI \citep{Gilli-2006}, and Ti \citep{Valenti-2005} and a systematic overabundance trends in Mn \citep{Gilli-2006}, Mg, and Al \citep[][]{Gilli-2006,Bensby-2005}. We emphasise the existence of a population of stars with a systematic underabundance in Si \citep{Valenti-2005} and verified that this is not a planet-host population. We do not know the origin of these differences. However, the differences are, in general, small and systematic, implying that, from a relative point of view, all studies agree well.

\section{Abundances in planet-hosts}
\label{sec:planet}

Although we study a relatively small number of planet-host stars ({68}), this number is sufficient to observe whether there are any discernible differences in the abundances of stars with and without planets. 

We note that five planet-host stars (HD4308, HD27894, HD111232, HD114729, and HD330075) are considered to belong to the thick disc. This means that approximately 7\% of the planet-host stars in our sample belong to the thick disc and that approximately 17\% of the thick disc stars are planet hosts. These values must be regarded as lower limits, since (some?, most?) planets may still be awaiting confirmation in the entire sample. {It is important to note that there was no bias in the choice of our sample towards or against thick-disc planet-host stars}.

\begin{figure*}[t!]
\centering
\includegraphics[trim=0cm 1.5cm 0cm 1cm,clip,width=17cm]{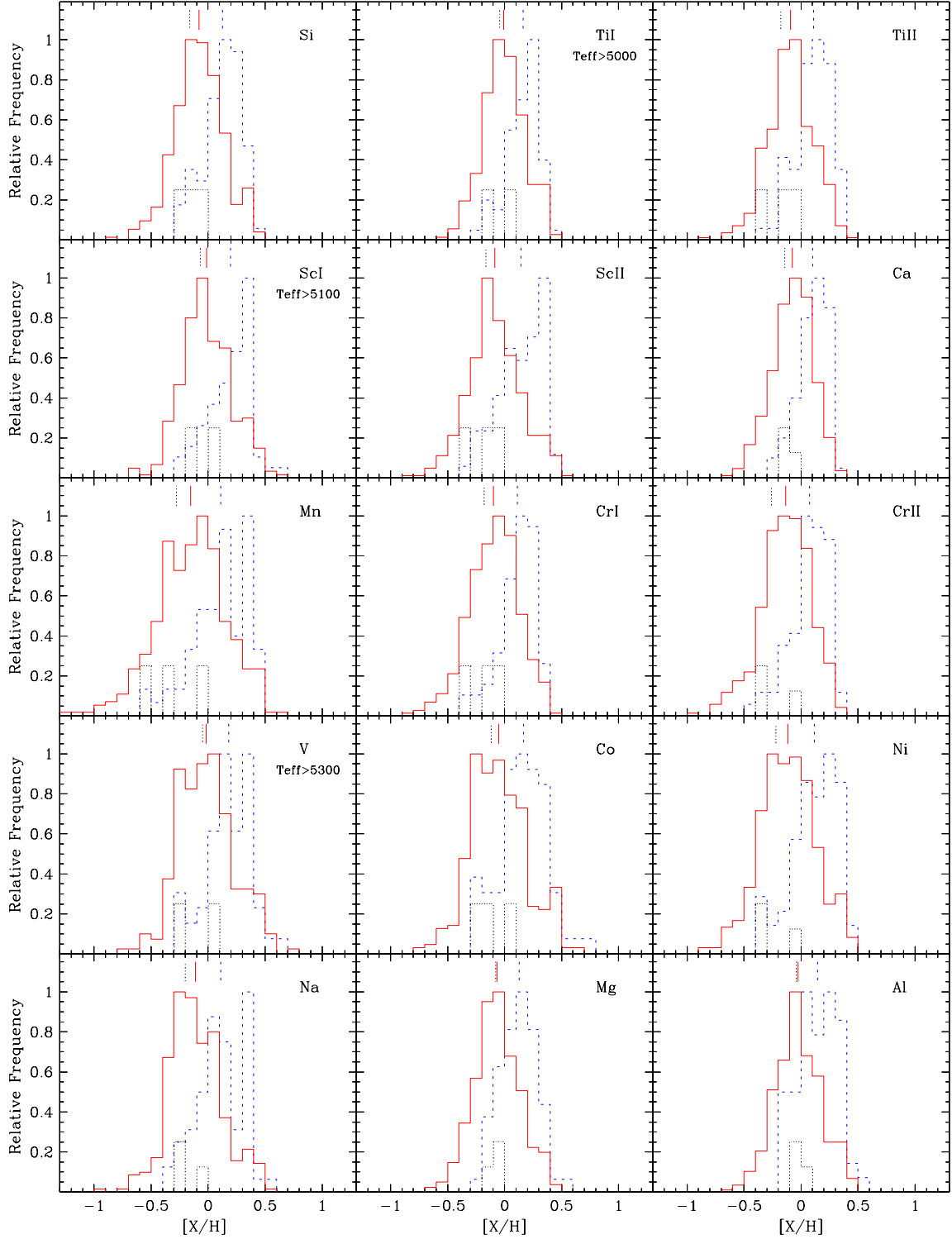}
\caption[depots]{[X/H] distribution of the different elements. The stars with and without planets are represented by a dashed blue, and solid red line respectively. The stars that exclusively host neptunians and super-earth planets are represented by a black dotted line. The mean value is pictured by a dashed (with planets), solid (without planets), and dotted (neptunian and super-earth hosts only) vertical lines on the top of each distribution. The neptunian and super-earth distribution was set smaller for clarity.}
\label{fig:distro}
\end{figure*}

\subsection{The [X/H] Distributions}

The [X/H] distributions of planet and non-planet hosts are depicted in Fig. \ref{fig:distro}. The stars with and without planets as well as the stars only with Neptunes and super-Earths\footnote{HD4308, HD40307, HD69830, HD47186, HD160691 and HD181433. The first three stars host neptunians and super-earth type planets only.} are represented by a dashed blue, solid red, and dotted black line, respectively. The vertical lines on the top of each histogram represent their average value.

We observe a clear metallicity excess for giant planet-host stars in all elements we study. These results are in good agreement with previous similar studies \citep[e.g.][]{Bodaghee-2003,Beirao-2005,Gilli-2006,Bond-2006,Takeda-2007}. Our results are also consistent with previous studies of iron abundances \citep[e.g.][]{Gonzalez-1998,Gonzalez-2001,Laws-2003,Santos-2001a,Santos-2003,Santos-2004b,Santos-2005a,Fischer-2005}, as expected. Interestingly, we also find that among thick disc members, stars with planets are more metal-rich  ($<$[Fe/H]$>$ = $-0.16$) than ``single'' stars ($<$[Fe/H]$>$ = $-0.33$).

Table \ref{table:avgabund} lists the average values of [X/H] along with their rms, {the number of stars used in their determination}, and the difference of averages between stars with and without planets. These differences range from 0.17 for Al to 0.27 for Mn. The last {four} columns list the average [X/H] ratio of Neptunian hosts and hosts with Neptunes and super-Earths only, {along with the number of stars used in their determination.}   

\begin{table*}[t!]
\centering
\caption[Average abundances for stars with and without planets ]{Average abundance values for stars with and without planets, along with their rms, {the number of stars used in their determination}, and the difference between the two groups. The last {four} columns list the average abundance values of neptunian hosts and hosts with neptunes and super-earths only, {along with the number of stars used in their determination.}}
\begin{tabular}{ l c c c c c c c c c c c}

\hline
\hline
Species \T & \multicolumn {3}{c}{Planet hosts} & \multicolumn {3}{c}{Non-planet hosts} & Difference   & \multicolumn {2}{c}{{Neptunian hosts}} & \multicolumn {2}{c}{Neptunian hosts only} \\
(X) \B & $\langle$[X/H]$\rangle$ & rms & {N} & $\langle$[X/H]$\rangle$ & rms & {N} & of Averages & $\langle$[X/H]$\rangle$  & {N} & $\langle$[X/H]$\rangle$ & {N} \\
\hline
Si &  0.13 &  0.17 & 68 & -0.08 &  0.21 & 383 &  0.21 &  0.08 & 6 & -0.16 & 3    \\
TiI &  0.16 &  0.14 & 62 & -0.01 &  0.18 & 321 &  0.17 &  0.12 & 4 & -0.05 & 2    \\
TiII &  0.11 &  0.15 & 68 & -0.09 &  0.20 & 383 &  0.20 &  0.04 & 6 & -0.18 & 3    \\
ScI &  0.20 &  0.19 & 61 & -0.01 &  0.22 & 288 &  0.21 &  0.14 & 4 & -0.07 & 2    \\
ScII &  0.14 &  0.19 & 68 & -0.09 &  0.23 & 383 &  0.23 &  0.12 & 6 & -0.17 & 3    \\
Ca &  0.10 &  0.14 & 68 & -0.08 &  0.18 & 383 &  0.18 &  0.05 & 6 & -0.14 & 3    \\
Mn &  0.11 &  0.24 & 68 & -0.15 &  0.31 & 383 &  0.27 &  0.05 & 6 & -0.28 & 3    \\
CrI &  0.11 &  0.17 & 68 & -0.10 &  0.22 & 383 &  0.21 &  0.06 & 6 & -0.18 & 3    \\
CrII &  0.07 &  0.17 & 68 & -0.14 &  0.22 & 383 &  0.21 & -0.01 & 6 & -0.26 & 3    \\
V &  0.18 &  0.20 & 56 & -0.02 &  0.25 & 243 &  0.20 &  0.15 & 4 & -0.05 & 2    \\
Co &  0.17 &  0.21 & 68 & -0.05 &  0.25 & 383 &  0.22 &  0.19 & 6 & -0.12 & 3    \\
Ni &  0.12 &  0.20 & 68 & -0.12 &  0.24 & 383 &  0.23 &  0.07 & 6 & -0.22 & 3    \\
Na &  0.11 &  0.20 & 68 & -0.11 &  0.23 & 383 &  0.22 &  0.10 & 6 & -0.20 & 3    \\
Mg &  0.13 &  0.16 & 68 & -0.07 &  0.20 & 383 &  0.19 &  0.12 & 6 & -0.08 & 3    \\
Al &  0.15 &  0.17 & 68 & -0.03 &  0.20 & 383 &  0.17 &  0.17 & 6 & -0.04 & 3    \\
Fe$^*$ &  0.10 &  0.17 & 68 & -0.12 &  0.23 & 383 &  0.22 &  0.03 & 6 & -0.24 & 3   \\
\hline
\multicolumn{5}{l}{* [Fe/H] values from \citet{Sousa-2008}.}
\label {table:avgabund}
\end{tabular}
\end{table*}

The difference in the average abundances of stars with and without giant planets is, on average, slightly above the values measured by \citep[][]{Takeda-2007} by 0.03\,dex. This difference is also above the results of \citet{Bond-2006} for all elements. Our results are, on average, similar to those of \citet{Bodaghee-2003} and \citet{Gilli-2006}. The dispersion in this differences for all elements is the smallest of all authors, which probably attests to the quality of our results.

It is interesting to see that in most histograms (except for Mg) the distributions of the abundances in planet-host stars are asymmetrical: there is an increase in [X/H] to a maximum value and shortly afterwards there is a significant cutoff in distribution. This cutoff is located at [X/H] $\sim$ 0.3 for Ca, and CrII, [X/H] $\sim$ 0.4 for Si, TiI, TiII, ScI, ScII, CrI, Ni, Na, Mg, and Al and [X/H] $\sim$ 0.5 for Mn, V, and Co. The cutoff might suggest that this is the metallicity limit of the solar neighbourhood stars  \citep[e.g.][]{Santos-2003}, since most of the planet hosts are at the high metallicity end of the sample.

Some planet-host distributions appear to be slightly bimodal (Mn, V, and Na). We cannot explain this effect which might be due only to small number statistics.

Regarding stars with only Neptunian planets (N=3), we observe that they have the lowest [X/H] mean value for each element. We note that there are only 3 stars with Neptunes, and 3 stars that are Neptunian and Jovian hosts. Therefore, any results regarding the Neptunian and super-Earth planet hosts must be seen as preliminary. Despite that, the values for the Neptunian and super-Earth-only hosts confirm that the probability of finding one of these planets may be even higher in metal-poor stars, and they appear to exhibit a different trend from the Jovian-host stars. Our results for stars hosting only neptunian-like planets agree with those of \citet{Udry-2006, Udry-2007b} and \citet{Sousa-2008} for [Fe/H].

\begin{figure}[t!]
\centering
\includegraphics[trim=0cm 1.5cm 0cm 1cm,clip,width=9cm]{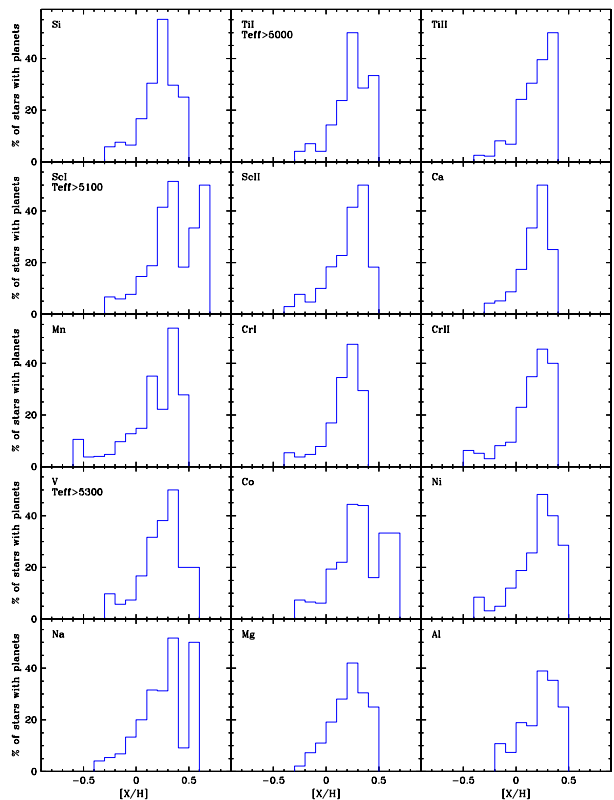}
\caption[depots]{Percentage of stars with planets as a function of [Fe/H]. The element label and the used cutoff temperature (when required) are located at the upper left corner of each plot.}
\label{fig:percent}
\end{figure}

Figure \ref{fig:percent} illustrates the [X/H] distribution in a different perspective: the histograms of the number of stars with planets relative to the total number of stars of each bin (0.1 dex). We observe that there is a general increase in the percentage of stars with planets for all elements, with increasing [X/H], after a low metallicity plateau. This observation agrees with the findings of other authors such as, for example, \citet{Santos-2001a, Santos-2004b}, and \citet{Fischer-2005} for [Fe/H].

\begin{figure*}[t!]
\centering
\includegraphics[trim=0cm 1.3cm 0cm 1.3cm,clip,width=17cm]{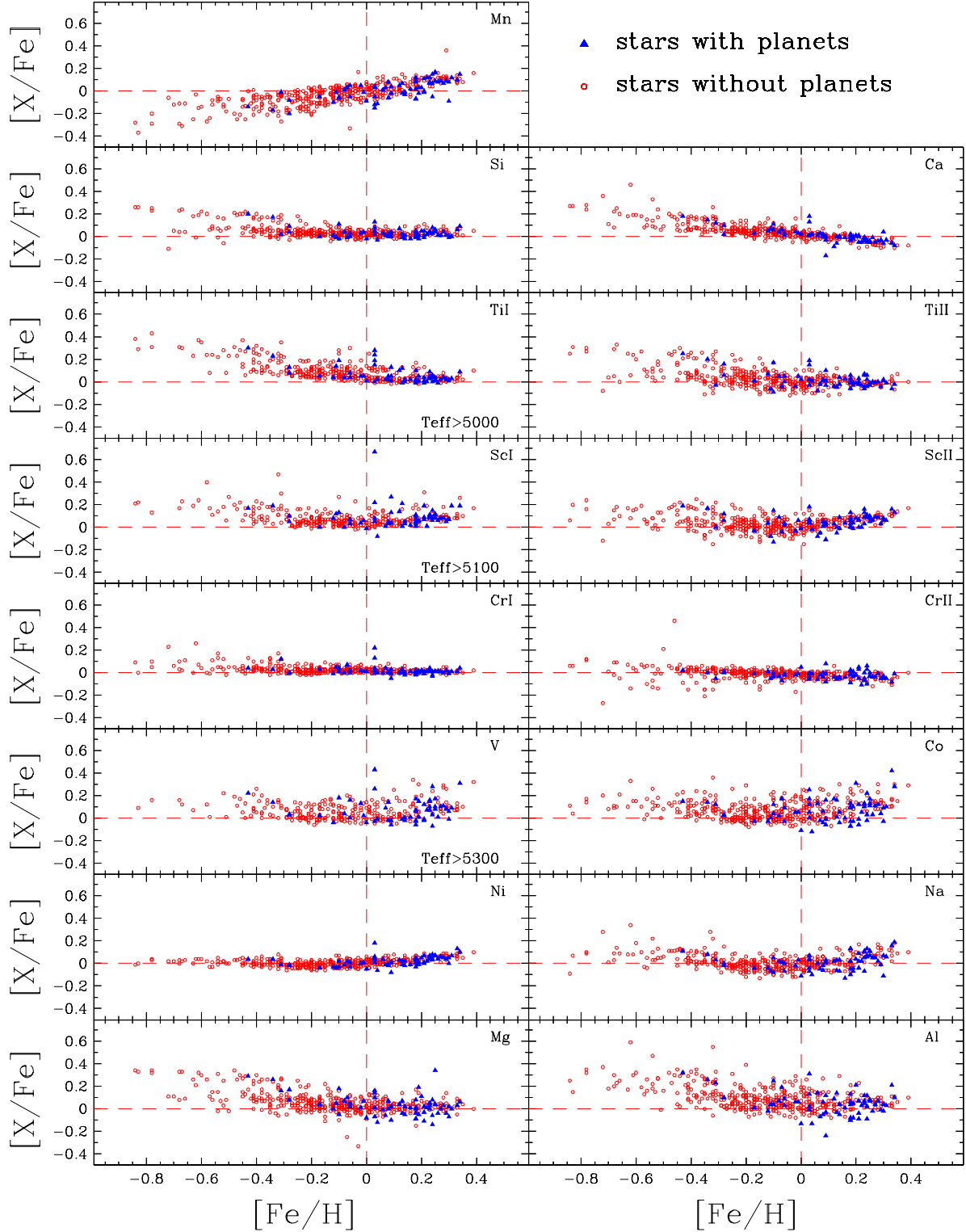}
\caption[abundance gfx]{[X/Fe] vs. [Fe/H] plots for the elements in study. The blue triangles and the red circles represent the stars with and without planets, respectively. The intersection of the dashed lines indicate the solar value. Each element is identified in the upper right corner of the respective plot and the cutoff temperature is indicated in the lower right corner when appropriate.} 
\label{fig:xfefeh1}
\end{figure*}

\subsection{The [X/Fe] versus the [Fe/H] plots: stars with and without planets}
\label{section:xfefeh}

\begin{table*}[t!]
\centering
\caption[]{Average abundance values for the stars from the thin disc, transition and thick disc populations with [Fe/H] $< 0$, along with their rms. The last two columns depict the difference of the average abundance and the KS probabilities between the thin and the thick disc populations.}
\begin{tabular}{ l r r r r r r c c}

\hline
\hline
Species \T & \multicolumn {2}{c}{Thin Disc} & \multicolumn {2}{c}{Intermediate} & \multicolumn {2}{c}{Thick Disc} & Difference & KS \\
(X) \B & $\langle$[X/Fe]$\rangle$ & rms & $\langle$[X/Fe]$\rangle$ & rms & $\langle$[X/Fe]$\rangle$ & rms & of Averages & probability \\
\hline
Si & 0.04 & 0.04 & 0.10 & 0.07 & 0.14 & 0.08 & 0.10 & 2.9e-09 \\
TiI & 0.10 & 0.07 & 0.19 & 0.10 & 0.22 & 0.11 & 0.12 & 6.7e-06 \\
TiII & 0.03 & 0.07 & 0.13 & 0.10 & 0.19 & 0.09 & 0.16 & 2.7e-11 \\
ScI & 0.07 & 0.06 & 0.14 & 0.11 & 0.15 & 0.07 & 0.08 & 1.5e-06 \\
ScII & 0.02 & 0.06 & 0.11 & 0.07 & 0.14 & 0.07 & 0.12  & 2.1e-11 \\
Ca & 0.06 & 0.06 & 0.12 & 0.12 & 0.14 & 0.07 & 0.08 & 8.9e-07 \\
Mn & -0.05 & 0.07 & -0.12 & 0.12 & -0.17 & 0.08 & 0.12 & 5.6e-10 \\
CrI & 0.03 & 0.04 & 0.04 & 0.06 & 0.03 & 0.03 & 0.00  & 6.5e-01 \\
CrII & -0.01 & 0.05 & 0.01 & 0.06 & 0.04 & 0.10 & 0.05  & 2.7e-05 \\
V & 0.05 & 0.06 & 0.10 & 0.07 & 0.11 & 0.08 & 0.06  & 3.1e-03 \\
Co & 0.05 & 0.08 & 0.11 & 0.09 & 0.10 & 0.06 & 0.05 & 8.3e-05 \\
Ni & -0.01 & 0.03 & 0.02 & 0.02 & 0.02 & 0.03 &0.03 & 2.6e-04 \\
Na & -0.00 & 0.06 & 0.06 & 0.09 & 0.06 & 0.06 & 0.06 & 6.1e-07 \\
Mg & 0.05 & 0.08 & 0.17 & 0.10 & 0.21 & 0.11 & 0.16 & 2.1e-09 \\
Al & 0.10 & 0.10 & 0.21 & 0.14 & 0.22 & 0.09 & 0.12  & 8.8e-06 \\
\hline
\label {table:avgabundTD}
\end{tabular}
\end{table*}

The [X/Fe] versus [Fe/H] plots are traditionally used to study the chemical evolution of the galaxy as well as to identify its different stellar populations (thin disc, thick disc, and halo), which have distinct abundance ratios \citep[see e.g.][]{Bensby-2003,Fuhrmann-2004}. In this section, we analyse them only to determine whether there are any differences in the abundances of stars with and without planets, for the same value of [Fe/H].

Figure \ref{fig:xfefeh1} shows the [X/Fe] values for the two samples as a function of [Fe/H]. 
The blue triangles and the red circles represent stars with and without planets, respectively. The red dashed lines target the solar value. The cutoff temperature adopted is indicated where appropriate (see Sect. \ref{sec:testing}). 

The plots do not show any clear differences between the stars with and without planets for a given [Fe/H] value. We also completed a Kolmogorov-Smirnov (KV) test of the [X/Fe] distributions for [Fe/H] $> 0$ without identifying any significant differences between the planet-host and the non-planet-host populations. The results are in general agreement with previous similar studies such as, for instance, \citet{Gonzalez-2001}, \citet{Takeda-2001},  \citet{Sadakane-2002}, \citet{Bodaghee-2003}, \citet{Beirao-2005}, \citet{Fischer-2005}, \citet{Gilli-2006}, and \citet{Takeda-2007}, which encourages  confidence in our results.

The reported cases of potential trends described by other authors such as, for instance, \citet{Sadakane-2002} for vanadium and cobalt, \citet{Bodaghee-2003} for Ti, Mn, V, and Co, and \citet{Gilli-2006} for V, Co, Mg, and Al, were not found. \citet{Robinson-2006} claimed to have found clear and significant overabundances of [Si/Fe] and [Ni/Fe]. 
\citet{Gonzalez-2007} also found that there were strong indications that Al and Si abundances are systematically smaller for the planet-host stars in the higher metallicity region, and that the Ti abundance exhibits the opposite trend, hinting that the abundances of Na, Mg, Sc and Ni might have some differences between planet-hosts and non-planet-hosts. We could not identify any of these trends clearly in our data.

To avoid systematic errors, we also completed the analysis only for stars with $T_{eff}=T_\odot\pm$300 K. These plots provided us with a more accurate picture of the trends and differences among the two groups of stars: in a differential analysis such as this, the closer the star temperature is to the solar temperature, the more accurate, on average, the derived abundance is. A KS test was also completed for the [X/Fe] distribution of this temperature-restricted population. We did not find any new trend.

Some outliers may however exist in our sample. In all cases, the abnormal abundances may be explained by their high dispersion in individual line abundances, by a high value of EP and/or RW slopes (see Sect. \ref{sec:testing}), or as being due to non-LTE systematic effects for the highest temperature stars.

We did not identify any abundance anomalies in the stars previously described in the literature: the underabundance of Ca in HD209100 derived by \citet{Bodaghee-2003} 
and the overabundance of Mg and Al in the host star HD168746 described by \citet{Sadakane-2002}, \citet{Laws-2003}, and \citet{Gilli-2006}  was not found.

\section{Thin and thick disc}
\label{sec:disc}

{In this section we perform} a detailed analysis of the [X/Fe] distributions of three different populations: thin disc, thick disc,  and the intermediate population, for [Fe/H] below the solar value. Above [Fe/H] $>0$, we were unable to identify any clear difference between thin and thick disc stars. 
The chosen criteria for selecting stars in each population and the necessary data for determining membership were described in Sect. \ref{sec:uvwdata}. We must note that we only have 6 thick disc stars with [Fe/H] above the solar value. HD27894 is the thick disc member with the highest metal content of approximately 0.2 dex.

\subsection {The [X/Fe] distributions}

\begin{figure*}[t!]
\centering
\includegraphics[trim=0cm 1.5cm 0cm 1cm,clip,width=17 cm]{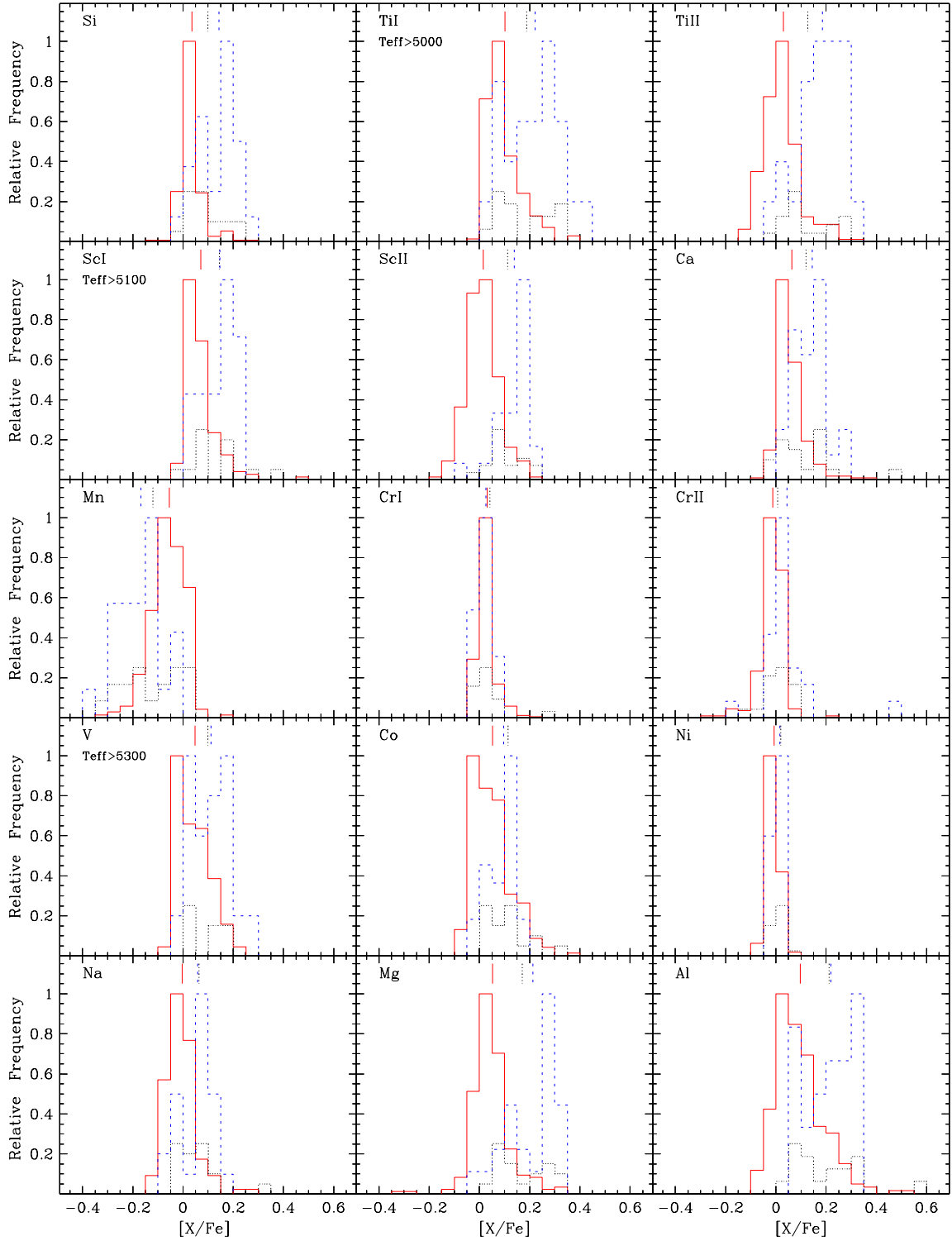}
\caption[]{[X/Fe] distributions of the species for [Fe/H]$<0$. Each element and ion is identified on the top left corner of the plots, along with the used cutoff temperature when relevant. The solid red, dotted black and dashed blue lines represent the thin, intermediate and thick disc populations, respectively. The dotted black histogram was set smaller for clarity. The mean value is depicted by the vertical lines above each distribution, corresponding to the respective populations.}
\label{fig:histxfeTD}
\end{figure*}

The [X/Fe] distributions for [Fe/H] $< 0$ are depicted in Fig. \ref{fig:histxfeTD}. Each element and ion is identified at the top left corner of the plots, along with the adopted cutoff temperature when relevant. The solid red, dotted black, and dashed blue lines represent the thin, intermediate, and thick disc populations, respectively. The mean value of each population is depicted by the vertical lines above the distributions.

Table \ref{table:avgabundTD} lists the average values of [X/Fe] for the three populations and their rms values. The last two columns depict the difference of the average abundance and the KS probabilities between the thin and thick disc populations.

From the analysis of the histograms and the associated table, we can observe that in the metallicity domain [Fe/H]$<$0, thick disc stars exhibit higher average [X/Fe] values for most of the species. However, for CrI, there is no difference between the two populations and for Co and Ni, this difference is small (0.03 and 0.04 dex respectively). We can also see that for all elements there is at least a partial superposition between the thin and the thick disc populations. The manganese histograms also display a different trend from the remaining elements: the values of thin and thick disc populations invert, and, this time, the thick disc has an underabundance relative to the thin disc.
The difference between both populations in this element is 0.12 dex.

The K-S probabilities imply that all the stars from the thick and the thin disc are indeed from separate groups, with the exception of CrI, which has a 65\% probability of belonging to the same population. We note that the results based on CrII data differ from those of CrI. However, CrII only exhibits two or three lines and, therefore, its results might be affected by more significant bias. The mean values of the transition population are situated between those of the other populations, but its average oscillates somehow for some elements.

It is difficult to conclude anything about the origin of this intermediate population. This is also valid for any statistical feature of the transition or the thick disc populations that might appear in the histograms: the data set is too small in size to enable reliable statistical measurements to be made. Nevertheless, we can observe that most thick disc histograms appear to be bimodal (Si, TiI, TiII, Ca, Mn, V, Co, Na, Mg, and Al): this may indicate that there are two populations in the thick disc with distinct [X/Fe] values. {Alternatively, the observed behaviour could be due to the fact that some stars from the thin disc may be misclassified as thick disc stars. The stars belonging to the lower [X/Fe] branch of the thick disc population have typically lower values of $P_{thick disc}/P_{thin disc}$ when compared to the upper-branch thick-disc stars (for Mn this tendency is opposite, as expected).}

With these results we confirm that the thin and thick disc populations are chemically and statistically distinct (at least in the region below solar metallicity). Besides the differences already discovered by \citet{Fuhrmann-1998} for Mg and by \citet{Bensby-2003} for Si, Ca, Ti and Al, we also found that Sc, Mn, Co and Na display the same separation between the aforementioned populations. The results of \citet{Bensby-2005} for Na differ from ours: for this element, they found that the thin-disc stars might be more abundant (have higher [X/Fe]) than thick-disc stars. We found the opposite trend. Regarding Ni, we can only say that there is a hint, also found by \citet{Bensby-2003}, that the thick-disc stars might be more abundant than thin-disc stars at [Fe/H] $<0$. 

\subsection{The [X/Fe] versus [Fe/H] plots: thin and thick disc populations}
\label{sec:xfefehTD}

\begin{figure*}[t!]
\centering
\includegraphics[trim=0cm 1.3cm 0cm 1.3cm,clip,width=17 cm]{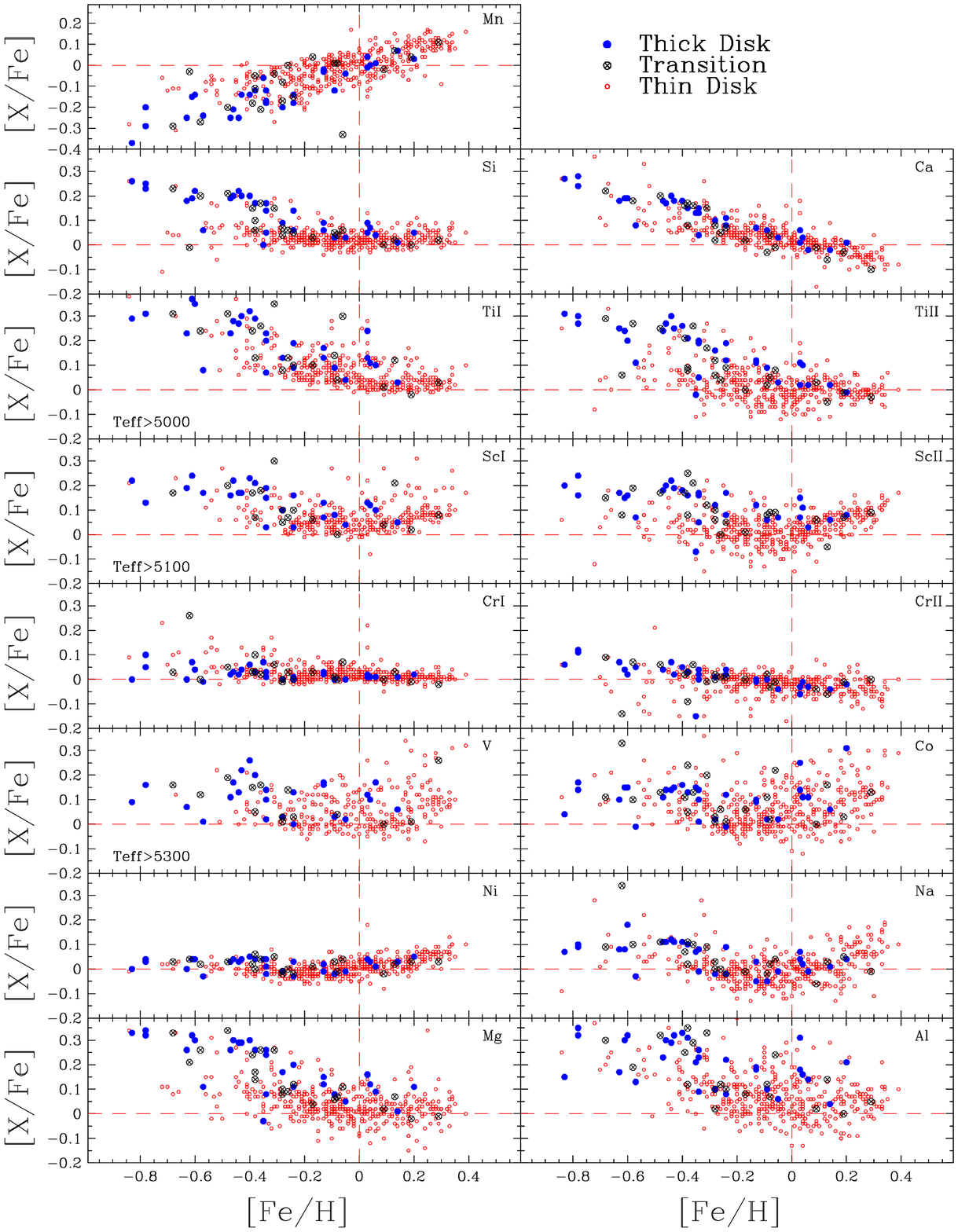}
\caption[abundance gfx]{[X/Fe] vs. [Fe/H] plots. The blue, black crossed and empty red circles represent the thick disc, transition and thin disc stars, respectively. The dashed lines mark the solar value. Each element is identified in the upper right corner of the respective plot and the cutoff temperature is indicated in the lower left corner when appropriate.}
\label{fig:xfefeh1TD}
\end{figure*}

\begin{figure*}[t!]
\centering
\includegraphics[trim=0cm 1.3cm 0cm 1.3cm,clip,width=17 cm]{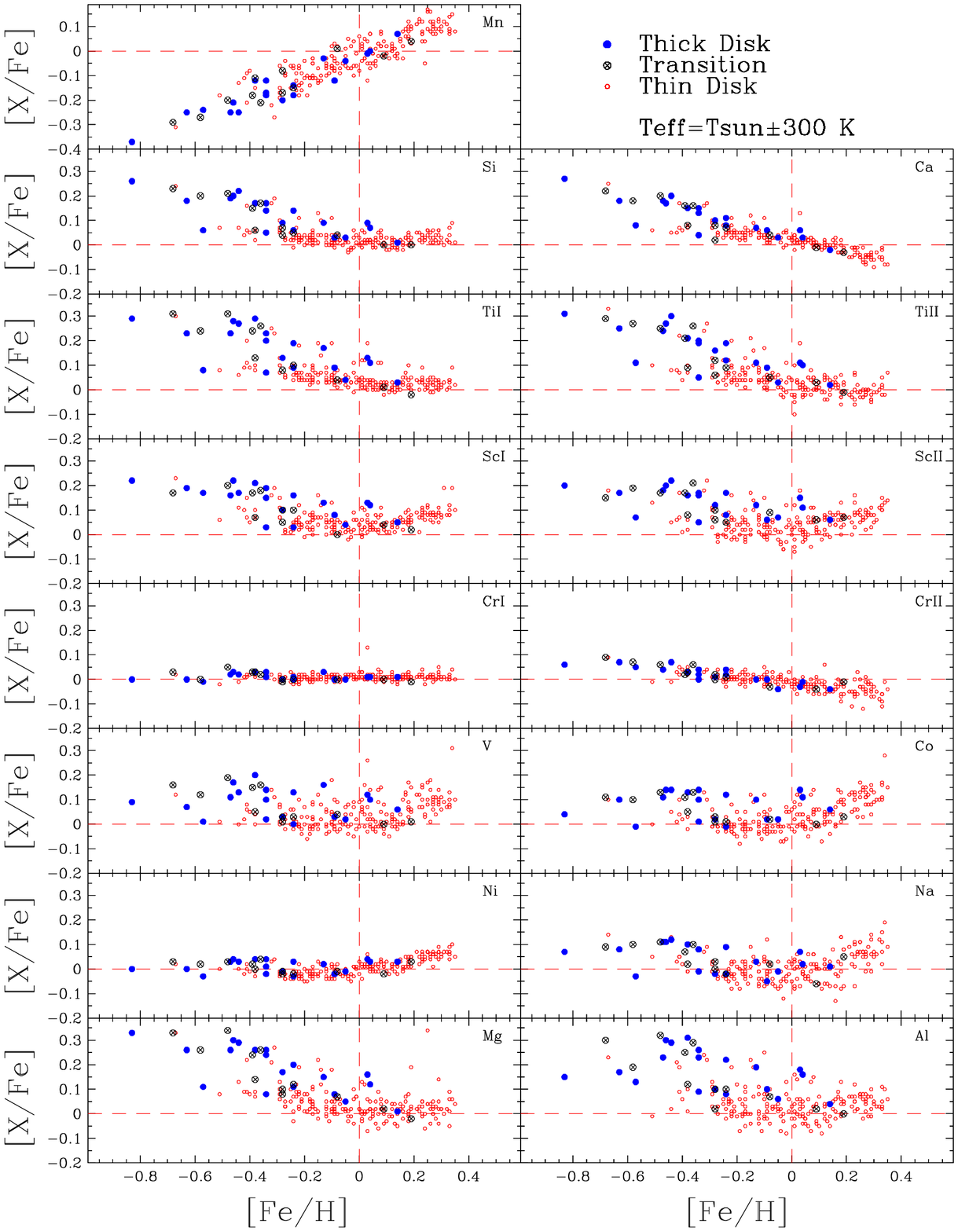}
\caption[abundance gfx for solar temperatures]{Same as Fig. \ref{fig:xfefeh1TD} but only with stars having $T_{eff}=T_\odot\pm$300 K.}
\label{fig:xfefeh2TD}
\end{figure*}

The [X/Fe] versus [Fe/H] plots are depicted in Figs. \ref{fig:xfefeh1TD} and \ref{fig:xfefeh2TD}. {The blue, black crossed, and red empty circles} represent the thick-disc, transition, and thin-disc stars, respectively. The dashed lines indicate the solar value. In Fig. \ref{fig:xfefeh2TD}, only stars with $T_{eff}\pm300 K$ are shown: this allows us to obtain a clearer picture of the trends in the two groups.

Following the results of the [X/Fe] distributions, in Fig. \ref{fig:histxfeTD}, we verify that, in the [X/Fe] versus [Fe/H] plots, the abundance difference between the thin and thick disc populations is not clearly defined. We can observe that, for [Fe/H] $<0$, the thin and the thick disc populations are mixed, and for a fixed value of [Fe/H], we cannot conclude that the populations are fully separated. Instead, for the $\alpha$ elements (Si, Ca, Ti, and Sc) and magnesium, the plots reveal a bifurcation into two different groups of stars, containing members from the three considered populations, that converges at [Fe/H] between -0.2 dex and the solar value. There are some hints that this might also be the case for V, Co, Na, and Al. 
These trends are clearer in Fig. \ref{fig:xfefeh2TD}. The slope in the branch with the higher [X/H] is always steeper than the slope of the lower branch. This means that the change in the [X/H] ratio of the higher branch is faster for [Fe/H] below the solar value. 

We did not identify any differences between the planet and non planet-hosts among the thick disc members for a given [Fe/H] value. {However, we found some indication that metal-poor ([Fe/H] $\leqslant-0.2$) planet-host stars originate preferentially in the thick disc, as also suggested by \citet{Haywood-2008}. Three of the five planet hosts in this metallicity domain are classified as thick disc stars, one as a transition object and the remainder as member of the thin disc. However, we note that we have only five stars in these conditions, and therefore, are unable to confirm reliably the validity of this result.}

\citet[][]{Bensby-2003} and \citet[][]{Bensby-2005} concluded that there is a clear separation between stars from the thick and thin disc populations.
No reference to the existence of the bifurcation discussed here was given by these authors.
We are still unable to ascertain its origin, although as discussed in the previous section, it may indicate that some thin-disc members have been misclassified.

\subsection{The [X/Fe] versus [Fe/H] plots: galactic chemical evolution}
\label{sec:galaxy}

It is widely accepted that the chemical evolution of the Galaxy is dominated by nucleosynthesis occurring in many generations of stars, with the exception of the lightest elements \citep{McWilliam-1997}. Here, we use the [X/Fe] versus [Fe/H] relations, in Figs. \ref{fig:xfefeh1TD} and \ref{fig:xfefeh2TD}, to study the chemical evolution of the
galaxy, assuming that [Fe/H] can be used as a time variable.
It is difficult to say whether the heavy scatter found in some plots are {cosmic (astrophysical) or due to errors}.

In these figures, we consider the stars with and without planets as one group, but we identify stars in the thick and thin-disc, as well {as in an} intermediate group (see Sect. \ref{sec:uvwdata}). The analysis is completed element by element and compared with other studies of chemical abundances between stars with and without planets, namely, \citet{Sadakane-2002},  \citet{Bodaghee-2003}, \citet{Beirao-2005}, \citet{Fischer-2005}, \citet{Gilli-2006}, \citet{Gonzalez-2007}, and \citet{Takeda-2007} (hereafter SAD, BOD, BEI, FV05, GIL, GZ07, and TAK, respectively), as well as other reference studies by \citet{Chen-2000}, \citet{Allende-2004}, and \citet{Bensby-2005} (hereafter CH00, AP04, and BEN, respectively). 

\subsection{The alpha elements}

It is believed that the alpha elements (Si, Ca, Ti, and Sc in this study) are produced mostly in massive stars that explode as Type II supernova (SNe II), although Type I supernova (SNe Ia) might also yield some of these elements \citep[e.g.][]{Thielemann-2002}. 

We can observe that for all elements in this group, there is a bifurcation in the value of [Fe/H] below solar. The upper branch descends from overabundances of $\sim$ 0.3 and $\sim$ 0.4 dex toward solar abundances at [Fe/H] around the solar value, while the lower branch has a much slower descent. The abundance then remains at solar value and at some point it {ascends, descends, or remains constant for [Fe/H] above the solar value}, depending on the element: this might be observational evidence that the $\alpha$ elements are produced in different amounts by different SN \citep{McWilliam-1997}. We should note, however, that titanium is overabundant relative to the other $\alpha$ elements at [Fe/H] below the solar value, and that its {slope} (in this metallicity region) is steeper than for calcium and silicon. The Sc abundances, for [Fe/H] $< 0$, seem to slowly decrease or remain constant until [Fe/H] reaches $\sim-0.3$ dex, where it has a downturn toward solar values.

\subsubsection{Silicon}

We can see that there exists a clear bifurcation in a region of [Fe/H] ranging from -0.85 to 0.0 dex. This can be seen more clearly in Fig. \ref{fig:xfefeh2TD}. The [Si/Fe] value of the upper branch decreases until [Fe/H] $\sim -0.2$ dex, before joining with the lower branch, that has a much gentler descent. Then, the [Si/Fe] ratio remains constant at solar value, with a slight upwardly trend starting at [Fe/H] $\sim +0.2$.

The results of other authors are, in general, similar to our own. However, the plateau described also in CH00 appears to start much earlier than in our case, and in BEN there seems to be no plateau. BOD, FV05, and GZ07 obtained slightly lower abundances ($\sim$ 0.05 dex) in the plateau region. The upward trend at [Fe/H] $>0$ is not evident in SAD, FV05, GIL, or TAK. The AP04 measurements are systematically higher than ours ( by more than $\sim +0.1$ dex) and the upward slope is far more significant than in our results. No author refers to the bifurcation observed here.

\subsubsection{Calcium}

Calcium shows an uniform downward trend for the entire metallicity range. It begins from abundance values of $\sim +0.3$ at [Fe/H] $\sim -0.85$ dex, and decreases to [Ca/Fe] values of $\sim -0.1$ at [Fe/H] $\sim +0.4$, equalling the solar value at solar [Fe/H]. The branches of the bifurcation are more difficult to observe and they seem to join at [Fe/H] $\sim -0.3$ dex. A slight plateau might exist for solar abundance values {in the metallicity region} between $-0.2$ and the solar value. This is more evident in Fig. \ref{fig:xfefeh2TD}.

The trends obtained for calcium are similar to most studies found in the literature, such as CH00, BOD, GIL, TAK and GZ07. However, the abundances obtained by SAD, AL4, and BEN do not continue to descend for [Fe/H]$>0$. The results of GIL and GZ07 present the same trend observed here, but the [Ca/Fe] values are systematically lower than ours. No author refers to the presence of a bifurcation.

\subsubsection{Titanium}

The upper branch of both TiI and TiII plots decrease until [Fe/H] reaches $\sim$ 0.0-0.2 dex, where it joins the lower branch. It then settles there for the remainder of the [Fe/H] range, at solar [Ti/Fe] value. Since the plots of Ti exhibit considerable scatter, it is easier to observe the trends in Fig. \ref{fig:xfefeh2TD}. We can see that the higher branch population has a [Ti/Fe] value between $\sim +0.35$ and $\sim 0.0$ dex, in a region of [Fe/H] ranging from $-0.85$ to $\sim 0.0$ dex.

A similar behaviour in the titanium trends is found in the studies of CH00, SAD, BOD, GIL, BEN, TAK, and GZ07. The plateau region in BOD, however, appears to be slightly below the solar value. The observed trend in FV05 has a continuously decreasing trend throughout the [Fe/H] range. The data from AP04 exhibit an upturn in the [Ti/Fe] values above solar metallicity. We did not detect the weak upturn in TiII reported by TAK. Again, no author refer to the observed bifurcation.

\subsubsection{Scandium}

There is considerable scatter in the points in Fig. \ref {fig:xfefeh1TD} for the Sc plot. The trend is easier to observe in Fig. \ref {fig:xfefeh2TD}. In the ScI panel, we observe a slow decrease in the abundance of both branches until they join at [Fe/H] $\sim -0.1$. It remains constant at the solar value in the region of $-0.1 \lesssim$ [Fe/H] $\lesssim +0.1$, and there is then an increase in the abundance for [Fe/H] $\gtrsim +0.1$ dex. The ScII plot is similar to that of ScI, but there appears to be no plateau: the [ScII/Fe] value decreases until the [Fe/H] reaches the solar value and then begins to increase toward the limit of the [Fe/H] range.

The results of BOD, GIL, and GZ07 differ slightly from ours. The results are similar for [Fe/H] $<0$, but then we observe an increase in the abundance with [Fe/H], which they do not observe. However, BOD suggests that there might be an increase for iron-rich stars, but this is inconclusive due to scattering. This increase was also noticed by GIL and GZ07, although it is elusive. Our figures suggest a much stronger trend. The results of SAD and TAK are similar to ours, although the ScI and ScII results of the {latter author} do not show the upturn or show a very weak one for metallicity values above the solar value. AP04 also reports the same trends, but their results have a systematic overabundance when compared to ours. No branch is found in the referred works.

\subsection{The iron peak elements}

It is believed that the iron peak elements are mostly created in SNe Ia explosions \citep{Thielemann-2002}. 
In our plots, we can observe that CrI and Ni (for [Fe/H] $<0$) follow iron in lockstep, which implies that they should have the same origin. On the other hand, vanadium and cobalt have trends more similar to the $\alpha$ elements. Manganese seems to mirror the behaviour of the alpha elements.

\subsubsection{Chromium}

In the CrI plot, the chromium abundances are constant around the solar value, as expected by the theory. The scatter is small when compared to the other elements. We note that the CrII plot has a weak, continuous downward trend but we only detect 2 or 3 lines per star, which means that the abundance determination is more affected by blending effects or to some other unknown systematic error.

The CrI results of CH00, SAD, BOD, BEN, and TAK are similar to ours. However, we observe a slight systematic underabundance in the results of BOD. The results of GIL for CrI suggest a downward trend similar to that in our CrII plot and their values have a systematic underabundance when compared with ours. Regarding our CrII plot, we find that it is different from the CrI plots of every author. The only work, besides ours, with a CrII determination is TAK: it shows a systematic upward trend throughout the metallicity range, in disagreement  with our results.

\subsubsection{Vanadium}

Vanadium has a behaviour similar to the alpha elements, which might indicate a common origin. However, we do not see a clear bifurcation, although there are some hints that it might exist (see Fig. \ref{fig:xfefeh2TD}). The [V/Fe] values slowly decline toward solar metallicity up to [Fe/H] $\sim -0.1$. The trend then becomes inverted and we can observe an increase toward the upper end of the [Fe/H] region. We can easily verify that there is a lot of scatter. We should note that the vanadium abundances could not have been determined if we had not removed the stars with $T_{eff}>5300$ K (see Sect. \ref{sec:testing}). Therefore, it is impossible to reach any important conclusion about the true trend or the origin of this particular element.

Our results for vanadium differ from those of CH00 and SAD, who reported that the vanadium abundances reflect that of Cr and Ni. However, the results of BOD, GIL and, TAK are similar to ours. Nevertheless, we note that almost every author reports high levels of scattering, which some authors propose is caused by NLTE effects (e.g. BOD, GIL).

\subsubsection{Cobalt}

Cobalt has a large scatter throughout the entire metallicity range. The abundance value decreases slowly until it reaches [Fe/H] $\sim -0.1$ dex. The [Co/Fe] ratio then remains constant around the solar value until [Fe/H] reaches the solar metallicity. Toward higher metallicities, there is a slow but clear increasing trend. In Fig \ref{fig:xfefeh2TD}, the abundance seems to be constant all the way to the solar value and a bifurcation similar to that found for the $\alpha$ elements might also be present. Cobalt has trends similar to both vanadium and the alpha elements. This element might be subjected to NLTE effects (see Sect. \ref{sec:testing}). Cobalt, as well as vanadium, are odd-Z nuclei but, despite that, they have trends similar to those of the  $\alpha$ elements.

The general trends observed by SAD, BOD, GIL and TAK are present in our plots. AP04 also exhibit a similar result. However, the upturn observed by AP04 for high [Fe/H] is steeper than those observed here. All abundance determinations, including ours, have considerable scatter. It is thus difficult to draw any strong conclusions from this analysis.

\subsubsection{Nickel}

The trends seen in Figs. \ref{fig:xfefeh1TD} and \ref{fig:xfefeh2TD} are all similar for this element. The [Ni/Fe] value remains roughly at solar value until [Fe/H] $\sim -0.3$ dex. It then has a small decrease, almost step-like, and remains in a plateau until [Fe/H] reaches the solar value. From this point on, it has two strong upturns at solar [Fe/H] and at $\sim +0.15$ dex. Between these values and from $\sim +0.15$ to the edge of the metallicity range, we can observe the existence of two small plateau with solar [Ni/Fe] ratio and [Ni/Fe] $\sim +0.05$, respectively. 
Nickel abundances have a low dispersion. We speculate that each one of these latter `steps' might correspond to events that initiated an increase in the production of Galactic Ni.

Our results are similar to those of BEN, although a little different in their fine details: they only report one upturn at solar metallicity. We note an approximate agreement with the results of CH00, SAD, FV05, and TAK, since they all present evidence of the same general trends (slow decrease up to the solar metallicity, followed by a weak upturn). Although BOD, GIL, and GZ07 report similar trends, their results have an abundance slightly below ours in the entire [Fe/H] range ($\sim -0.05$ dex). The same trends are also observed by AP04 but their abundance values are higher than most results and their slopes are also steeper.

\subsubsection{Manganese}

In contrast to all other elements, we observe a uniformly increasing trend for [Mn/Fe] values from $\sim -0.4$ to 0.1 dex throughout the entire [Fe/H] range. In Figs. \ref{fig:xfefeh1TD} and \ref{fig:xfefeh2TD}, we have the impression that two different branches exist for [Fe/H] values below $\sim$ 0.1 dex. The manganese underabundance in this region seems to mirror the $\alpha$-element overabundance. 

There is good agreement between our results and those of SAD and BOD. However, they do not find any hint of the existence of two separate branches. 
The GIL and TAK results differ from ours: they show a plateau for [Fe/H] below the solar value.

\subsection{The light metals: Mg, Na and Al}

Sodium and aluminium are understood to be created mostly in the cores of massive stars \citep{Chen-2000}, that later become SNe II. We can observe that the trends of Na and Al are similar to those of the alpha elements: we might even consider them to be mild alpha elements, even if we take into account the fact that they have an odd number of protons \citep{McWilliam-1997}. 
It is still difficult to differentiate the sources of Na and Al. Magnesium is proposed to be created only in {SNe\,II}, and can be considered a member of the alpha elements.

\subsubsection{Magnesium}

The [Mg/Fe] branches exhibit a decreasing trend until [Fe/H] $\sim -0.1$, although the upper branch has a much steeper descent. The branches then join and the abundance remains constant at solar value. We observe considerable scatter in the Mg plots,due to the fact that we used only 2 or 3 lines per star in the determination of its abundance. Magnesium is an interesting element: it is predicted that it can only be created in SNe II explosions \citep{Chen-2000}, but we can see that there is no continuous decline in magnesium as we would expect in this case \citep[][]{Chen-2000,Bensby-2003}. Therefore, a contribution of SNe\,Ia or other unknown factors may be causing the enrichment of Mg in the solar neighbourhood. 

The obtained trends are similar to those measured by SAD, AP04, BEI, BEN, GIL, and GZ07, but none of these authors refers to a bifurcation below solar [Fe/H] value. CH00 detect a similar trend but their plateau starts much earlier, at $-0.3$ dex, and the abundance value of this plateau for the entire range of [Fe/H] is higher than ours by 0.1\,dex. TAK, on the other hand, have a systematic underabundance in the same plateau above solar metallicity when compared with our results.

\subsubsection{Sodium}

Sodium abundance values slowly decrease until they reach the solar value at [Fe/H] $\sim$ -0.2 dex. They then remain constant for a metallicity range from [Fe/H] $\sim -0.2$ to $\sim -0.1$ dex. The abundance has an increasing trend for [Fe/H] $\gtrsim$ +0.2. There is considerable scatter in Fig. \ref{fig:xfefeh1TD}. It is easier to follow the trends in Fig. \ref{fig:xfefeh2TD}. There is also a slight hint in Fig. \ref{fig:xfefeh2TD} that there might be a bifurcation below solar [Fe/H]. The trend of Na is similar to the alpha-element trends, but the overabundance values at negative [Fe/H] are not so pronounced. 

A similar behaviour was found by SAD as well as by BEI, FV05, GIL, BEN, TAK, and GZ07. CH00 reported a constant abundance value throughout the entire [Fe/H] range which disagrees with our results. No author mentions the existence of a plateau in the intermediate [Fe/H] region. No bifurcation is reported in any paper.

\subsubsection{Aluminium}

The trends for aluminium are very similar to those of the alpha elements. One might say, at least from an observational point of view, that aluminium is an alpha element. Its abundance decreases until [Fe/H] $\sim$ -0.2 dex. It then remains at constant abundance ($\sim$$+$0.05 dex) until [Fe/H] reaches +0.1 dex. Afterwards, there is an upturn followed by a plateau with [Al/Fe] $\sim +0.05$ in the [Fe/H] region between 0.2 and 0.4 dex. The trends are easier to see in Fig. \ref{fig:xfefeh2TD}, where we can also see the presence of a possible bifurcation similar to the one in the Mg plot. Again, we observe more significant scatter than for sodium. 

Our results are similar to those obtained by BEI, GIL, and BEN. CH00 and SAD obtained different results from ours: the abundance remains constant at solar value and for CH00, it undergoes an increase toward higher metallicities at which it reaches $\sim$$-$0.3 dex. 

\section{Concluding remarks}
\label{sec:conclusion}

We derived the abundances of 12 elements (Si, Ca, Sc, Ti, V, Cr, Mn, Co, Ni, Na, Mg, and Al) in a sample of 451 field dwarfs from the HARPS ``high precision'' GTO planet search program, of which {68} planets are known to harbour planetary companions. The spectroscopic parameters used in this study were derived by \citet{Sousa-2008}.

We compared our results with those presented in other works to ensure consistency and reliability in our analysis. The [X/H] distributions as well as the plots of the abundance ratios [X/Fe] versus [Fe/H] were analysed in comparing the trends of stars with and without planets. 

We derived the galactic space-velocity components and used them to {identify kinematically} the origin of the stars of our sample, according to the method developed by \citet[][]{Bensby-2003} and \citet[][]{Bensby-2005}. We found that the sample consists of 400 stars from the thin disc, 29 stars from the thick disc, and 21 from a transition population that exhibits characteristics in common with both populations. One star is from the halo.

We confirmed that an overabundance in giant-planet {host-stars} is clear for all the studied elements. This may agree with previous suggestions that the efficiency of planetary formation correlates strongly with the metal content of the host star. In other words, the probability of finding a star with giant planets is higher in stars with high [X/H] ratios. The plot of the percentage of stars with planets as a function of the [X/H] ratio also provides us with this information. This dependence is also predicted by studies of core-accretion modelling \citep[see e.g.][]{Ida-2004b,Benz-2006}.
We also confirmed that stars hosting only neptunian-like planets may be easier to detect around stars with similar or even lower metallicity than non-planet hosts. This suggests that lower mass planets may have a different kind of formation than Jovian planets, which is also predicted by the aforementioned core-accretion studies of planetary formation. 

We do not find any evidence of differences between stars with and without planets for the same [Fe/H] in the plots of [X/Fe] versus [Fe/H]. This is in general agreement with previous studies. The stars that harbour planetary companions simply appear to be in the high metallicity tail of the distribution, following the same trends as the stars without planets. This result favours a primordial origin as the preferred cause of the observed high metallicities in planet-host stars. 

We found a clear distinction between thick disc and thin disc populations, for [Fe/H] $<$ 0, when we analysed the [X/Fe] distributions. This extends to all elements except chromium and nickel, and confirms and extends the general results found by \citet{Fuhrmann-1998}, \citet{Bensby-2003}, and \citet{Bensby-2005}. 

Interestingly, however, we observed that in the plots of [X/Fe] versus [Fe/H] the populations of the thin and thick disc are mixed for the same [Fe/H] throughout the entire metallicity range. We were unable to identify any separation between the populations as in \citet{Fuhrmann-2004} or \citet{Bensby-2005}. Instead, the plots of the alpha elements and Mg reveal, for [Fe/H] $<0$, a bifurcation of two different groups of stars, both with stars from the three populations, that converge at [Fe/H] close to the solar value. This bifurcation might also exist for V, Co, Na, and Al, but it is unclear due to scatter. In some plots (Si, Ca, Sc, Ti, Mg and Al), the two populations of this biffurcation seem to have different evolutionary dynamics: the one with lower abundances (dominated by {thin-disc} stars) has a shallow slope (if any), while the higher abundance population (dominated by {thick-disc} objects) has a far steeper slope, implying a more dynamic evolution. 

In general, the trends that we found agree with the literature references, with the exception of the aforementioned bifurcation. Nevertheless, we note that the traditional separations between alpha elements, iron peak elements, and others might need to be reformulated as the elements tend to have different trends, at least in the metallicity range of our study. Our results suggest that the origin of these elements is more complex than predicted by theory. Some elements exhibit a considerable scatter (V, Co, Na, Al), although we were unable to ascertain whether this scatter {is cosmic in origin or due to errors in the analysis}.

\begin{acknowledgements}
V.N. acknowledges the support given by Funda\c{c}\~{a}o Calouste Gulbenkian (Portugal).
N.C.S. would like to thank the support from Funda\c{c}\~ao para a Ci\^encia
e a Tecnologia, Portugal, through programme Ci\^encia\,2007
(C2007-CAUP-FCT/136/2006).
We would like to thank the HARPS team for having provided us with the data for this paper.
\end{acknowledgements}

\bibliographystyle{aa}
\bibliography{mylib.bib}

\end{document}